\documentclass[aps,reprint,twocolumns,longbibliography]{revtex4-1}
\usepackage{graphicx}
\usepackage{color}
\usepackage{upgreek}
\usepackage{graphics}
\usepackage{enumitem}
\usepackage{amsmath}
\usepackage{float}
\usepackage{siunitx}
\usepackage[table]{xcolor} 

\begin{document}

\title{Extracting quantitative biological information from brightfield cell images using deep learning}

\author{Saga Helgadottir$^{1,*}$}

\author{Benjamin Midtvedt$^{1,*}$}

\author{Jes\'us Pineda$^{1,*}$}

\author{Alan Sabirsh$^{2}$}

\author{Caroline B. Adiels$^{1}$}

\author{Stefano Romeo$^{3,4,5}$}

\author{Daniel Midtvedt$^{1}$}

\author{Giovanni Volpe$^{1}$}

\affiliation{$^{*}$ These authors contributed equally.} 
\affiliation{$^{1}$Department of Physics, University of Gothenburg, Sweden}

\affiliation{$^{2}$Advanced Drug Delivery, Pharmaceutical Sciences, R\&D, AstraZeneca, Gothenburg, Sweden}

\affiliation{$^{3}$Department of Molecular and Clinical Medicine, Institute of Medicine, Sahlgrenska Academy, Wallenberg Laboratory, University of Gothenburg, Sweden}

\affiliation{$^{4}$Department of Cardiology, Sahlgrenska University Hospital, Sweden}

\affiliation{$^{5}$Clinical Nutrition Unit, Department of Medical and Surgical Sciences, University Magna Graecia, Italy}

\date{\today}

\begin{abstract}
Quantitative analysis of cell structures is essential for biomedical and pharmaceutical research.
The standard imaging approach relies on fluorescence microscopy, where cell structures of interest are labeled by chemical staining techniques.
However, these techniques are often invasive and sometimes even toxic to the cells, in addition to being time-consuming, labor-intensive, and expensive. 
Here, we introduce an alternative deep-learning-powered approach based on the analysis of brightfield images by a conditional generative adversarial neural network (cGAN).
We show that this approach can extract information from the brightfield images to generate virtually-stained images, which can be used in subsequent downstream quantitative analyses of cell structures.
Specifically, we train a cGAN to virtually stain lipid droplets, cytoplasm, and nuclei using brightfield images of human stem-cell-derived fat cells (adipocytes), which are of particular interest for nanomedicine and vaccine development.
Subsequently, we use these virtually-stained images to extract quantitative measures about these cell structures. 
Generating virtually-stained fluorescence images is less invasive, less expensive, and more reproducible than standard chemical staining; furthermore, it frees up the fluorescence microscopy channels for other analytical probes, thus increasing the amount of information that can be extracted from each cell.
\end{abstract}

\maketitle

\begin{figure*}[t!]
    \centering
    \includegraphics[width=1\textwidth]{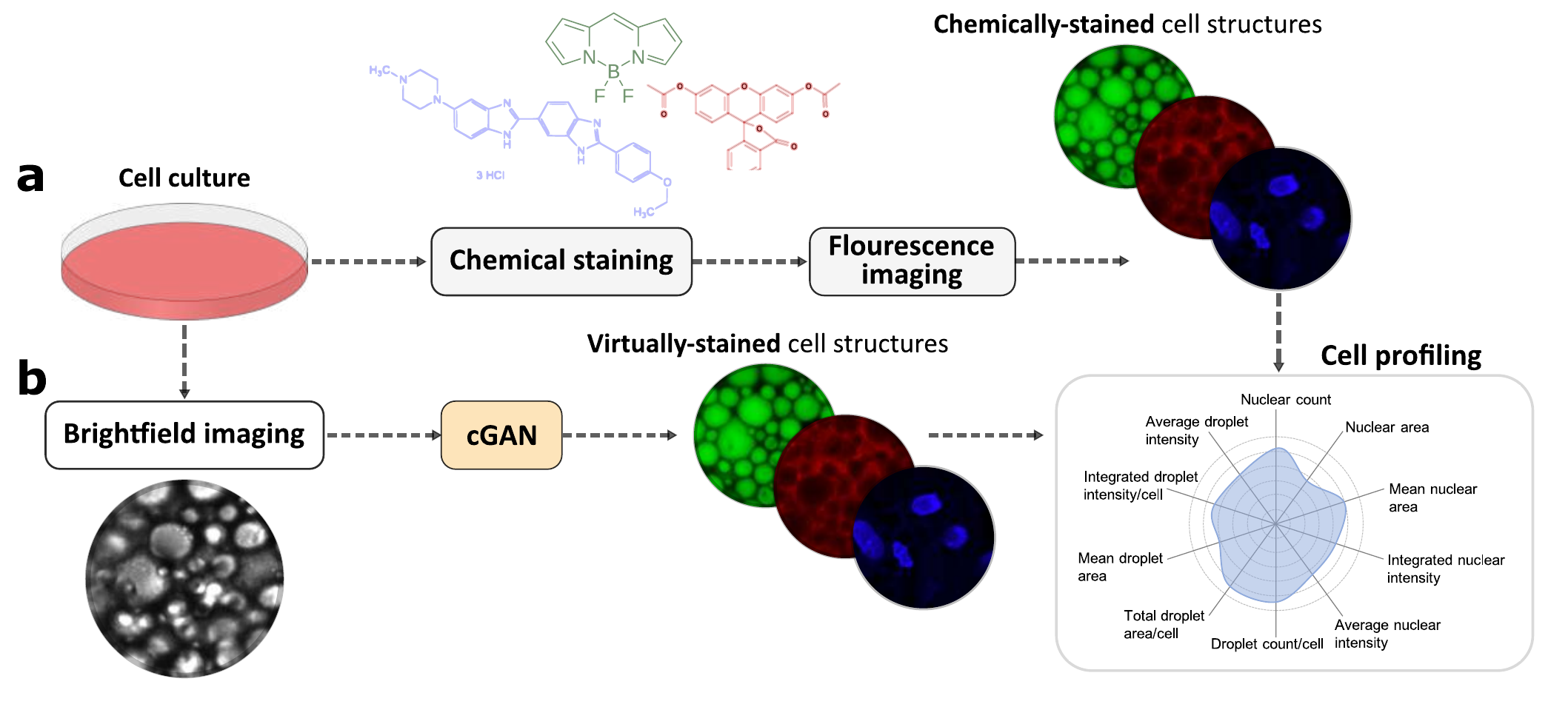}
    \caption{
    {\bf From cell cultures to quantitative biological information.} 
    {\bf a} The standard workflow entails chemically staining the cell structures of interest, imaging them using fluorescence microscopy (in multiple light channels), and, finally, using these fluorescence images to retrieve quantitative biologically-relevant measures about the cell structures of interest.
    {\bf b} The deep-learning-powered approach we propose replaces the chemical-staining and fluorescence microscopy with a conditional generative adversarial neural network (cGAN) that uses brightfield images to generate virtual fluorescence-stained images.}
    \label{fig:1}
\end{figure*}

Biomedical and pharmaceutical research often relies on the quantitative analysis of cell structures.
For example, changes in the morphological properties of cell structures are used to monitor the physiological state of a cell culture \cite{rizzuto1995chimeric},
to identify abnormalities \cite{bjorklund2006identification}, 
and to determine the uptake and toxicity of drugs \cite{kepp2011cell}.
The standard workflow is shown in Figure~\ref{fig:1}a: the cell structures of interest are chemically stained using fluorescence staining techniques; fluorescence images are acquired; and, finally, these images are analyzed to retrieve quantitative measures about the cell structures of interest.
One key advantage is that multiple fluorescence images of the same cell culture can be acquired in parallel using the appropriate combination of chemical dyes and light filters, with the resulting images containing information about different cell structures.

However, fluorescence cell imaging has significant drawbacks. 
First, it requires a fluorescence microscope equipped with appropriate filters that match the spectral profiles of the dyes. Besides the complexity of the optical setup, usually only one dye is imaged at each specific wavelength, limiting the combination of dyes and cell structures that can be imaged in a single experiment. 
Second, the staining of the cell structures is typically achieved by adding chemical fluorescence dyes to a cell sample, which is an invasive (due to the required culture media exchange and dye uptake \cite{lulevich2009cell}) and sometimes even toxic process \cite{alford2009toxicity}.
Third, phototoxicity and photobleaching can also occur while acquiring the fluorescence images, which results in a trade-off between data quality, time scales available for live-cell imaging (duration and speed), and cell health \cite{ounkomol2018label}.
Furthermore, a cell-permeable form of some dyes enters a cell, and then reacts to form a stable and impermeable reaction product that is transferred to daughter cells; as a consequence, the dye intensity dilutes at every cell division and is eventually lost. 
Fourth, fluorescence staining techniques are often expensive, time-consuming and labor-intensive, as they may require long protocol optimizations (e.g., dye concentration, incubation and washing times have to be optimized for each cell type and dye). Also, care has to be taken when choosing multiple dye partners to avoid spectral bleed-through \cite{zimmermann2005spectral}.
All these drawbacks aggravate, or hinder completely, the collection of reliable and long-term longitudinal data on the same population, such as when studying  cell behavior or drug uptake over time. 
Therefore, there is an interest in extracting the same information using cheaper, non-invasive methods.
In particular, it would be desirable to replace fluorescence images with brightfield images, which are much easier to acquire and do not require specialized sample preparation, eliminating concerns about the toxicity of the fluorescence dyes or damage related to the staining and imaging procedures.
However, while brightfield images do provide some information about cellular organization, they lack the clear contrast of fluorescence images, which limits their use in subsequent downstream quantitative analyses.

Recently, the use of deep learning has been proposed as a way to create images of virtually-stained cell structures, thus mitigating the inherent problems associated with conventional chemical staining.  
These proposals come in the wake of the deep learning revolution \cite{Lecun2015DeepLearning, cichos2020machine}, where convolutional neural networks have been widely used to analyze images, e.g., for microscopy \cite{barbastathis2019use} and particle tracking \cite{Hannel2018Machine-learningParticles, Newby2018Convolutional3D, Helgadottir2019DigitalLearning,  midtvedt2020quantitative}.
Virtually stained images have been created from images acquired with various imaging modalities.
For example, virtual staining of cells and histopathology slides has been achieved using quantitative phase imaging \cite{rivenson2019phasestain, zhang2020digital}, autofluorescence imaging \cite{rivenson2019virtual}, and holographic microscopy \cite{nygate2020holographic}.
Furthermore, more recent work suggests that the information required to reproduce different stainings is in fact available within brightfield images \cite{ounkomol2018label, liu2020global, li2020deep}.

Here, we propose a deep-learning-based approach to extract quantitative biological information from brightfield microscopy. 
A high-level description of the proposed workflow is shown in Figure~\ref{fig:1}b. 
Specifically, we train a conditional generative adversarial neural network (cGAN) to use a stack of brightfield images of human stem-cell-derived adipocytes to generate virtual fluorescence-stained images of their lipid droplets, cytoplasm, and nuclei.
Subsequently, we demonstrate that these virtually-stained images can be successfully employed to extract a series of quantitative biologically-relevant measures in a downstream cell-profiling analysis.
In order to make this deep-learning-powered approach readily available for other users, we provide a Python software package, which can be easily personalized and optimized for specific virtual-staining and cell-profiling applications.

\section*{Results}

\subsection*{Adipocyte cell culture, imaging, and cell profiling}

Adipocytes, or fat cells, are the primary cell type composing adipose tissue. 
They store energy in the form of lipids, mainly triglycerides, in organelles called lipid droplets. 
Adipocyte cell cultures are commonly employed to study how the adipocyte metabolic profile responds to therapies for metabolic diseases such as diabetes and non-alcoholic fatty liver disease
\cite{buzzetti2016multiple}.
They are also important therapeutically as they are present in the subcutaneous skin layers, and many relatively complex therapeutics, such as nanomedicines, vaccines or biologicals, are dosed using subcutaneous injections. 
For example, in the case of nanomedicines and vaccines containing mRNA, the adipocytes are important for creating the active therapeutic protein product \cite{blakney2020effect}.

Human adipose sampling, stem-cell isolation and subsequent cellular differentiation are described in detail elsewhere \cite{bartesaghi2015thermogenic}. 
Briefly, to remove mature adipocytes and isolate stem cells, adipose biopsies are minced, digested, filtered, and centrifuged. 
For differentiation into adipocytes, 90\% confluent stem-cell cultures are treated with DMEM/F12 containing 3\% fetal calf serum (Gold; PAA) and supplemented with $100\,{\rm nM}$ dexamethasone (Sigma), $500\,{\rm \upmu M}$ 3-isobutyl-1-methyxanthine (Sigma), $0.85\,{\rm \upmu M}$ insulin, and $5\,{\rm nM}$ triiodothyronine (Sigma). Media are changed every other day during proliferation and differentiation, until the cells are fully differentiated (day 32).

The mature adipocyte cultures, fixed using 4\% paraformaldehyde, are chemically stained to label lipid droplets (Bodipy, green fluorescent), cell cytoplasm (Cell Tracker Deep Red, red fluorescent), and nuclei (Hoechst 33342, blue fluorescent). All fluorescent reagents are from Thermo Fisher Scientific and are used according to the manufacturer's instructions.

The cell cultures are imaged using a robotic confocal microscope (Yokogawa CV7000) equipped with a $60\times$ water-immersion objective (Olympus, UPLSAPO 60XW, NA=1.2) and a 16-bit camera (Andor Zyla 5.5).
Illumination correction is applied during acquisition so that the fluorescence intensities are consistent over the field of view. In each well, brightfield and fluorescence images are captured for 12 non-overlapping fields of view ($280\,{\rm \upmu m} \times 230\,{\rm \upmu m}$, $2560 \times 2160$ pixels), for a total of 96 fields of view.
For each field of view, a set of four images (one brightfield image and three fluorescence images for lipid droplets, cytoplasm, and nuclei) is acquired at 7 different $z$-positions separated by $1\,{\rm \upmu m}$.
Subsequently, the fluorescence images at different $z$-positions are projected onto a single image using a maximum intensity projection to create a single fluorescence image per field of view and fluorescence channel.

Using the maximum intensity projections of the confocal fluorescence images, semi-quantitative phenotypic data is extracted from cell structures using the open-source cytometric image analysis software CellProfiler (https://cellprofiler.org, version 4.07 \cite{mcquin2018cellprofiler}) and a custom-made analysis pipeline (the analysis pipelines are available in the supplementary information \cite{VirtualStainingRepo}). Measured parameters include object numbers (nuclei, cells, lipid droplets), morphological characteristics (areas), and intensity data. 

\subsection*{Neural network architecture}

Neural networks are one of the most successful tools for machine learning \cite{Lecun2015DeepLearning, mehlig2019artificial}. 
They consist of a series of layers of interconnected artificial neurons. 
These artificial neurons are simple computational units that, when appropriately trained, output increasingly meaningful representations of the input data leading to the sought-after result. Depending on the problem, the architecture of the neural network varies. In particular, generative adversarial networks (GANs) \cite{NIPS2014_5ca3e9b1} have been shown to perform well in image-to-image transformation tasks, including recently to realize virtual stainings \cite{rivenson2019phasestain, rivenson2019virtual, zhang2020digital, nygate2020holographic, li2020deep}. 
A GAN consists of two networks \cite{NIPS2014_5ca3e9b1}: a \emph{generator}, which generates images, and a \emph{discriminator}, which discriminates whether images are real or created by the generator. 
The \emph{adversarial} aspect refers to the fact that these two networks compete against each other: 
during the training, the generator progressively becomes better at generating synthetic images that can fool the discriminator, while the discriminator becomes better at discriminating real from synthetic images.

\begin{figure*}[t!]
    \centering
    \includegraphics[width=1\textwidth]{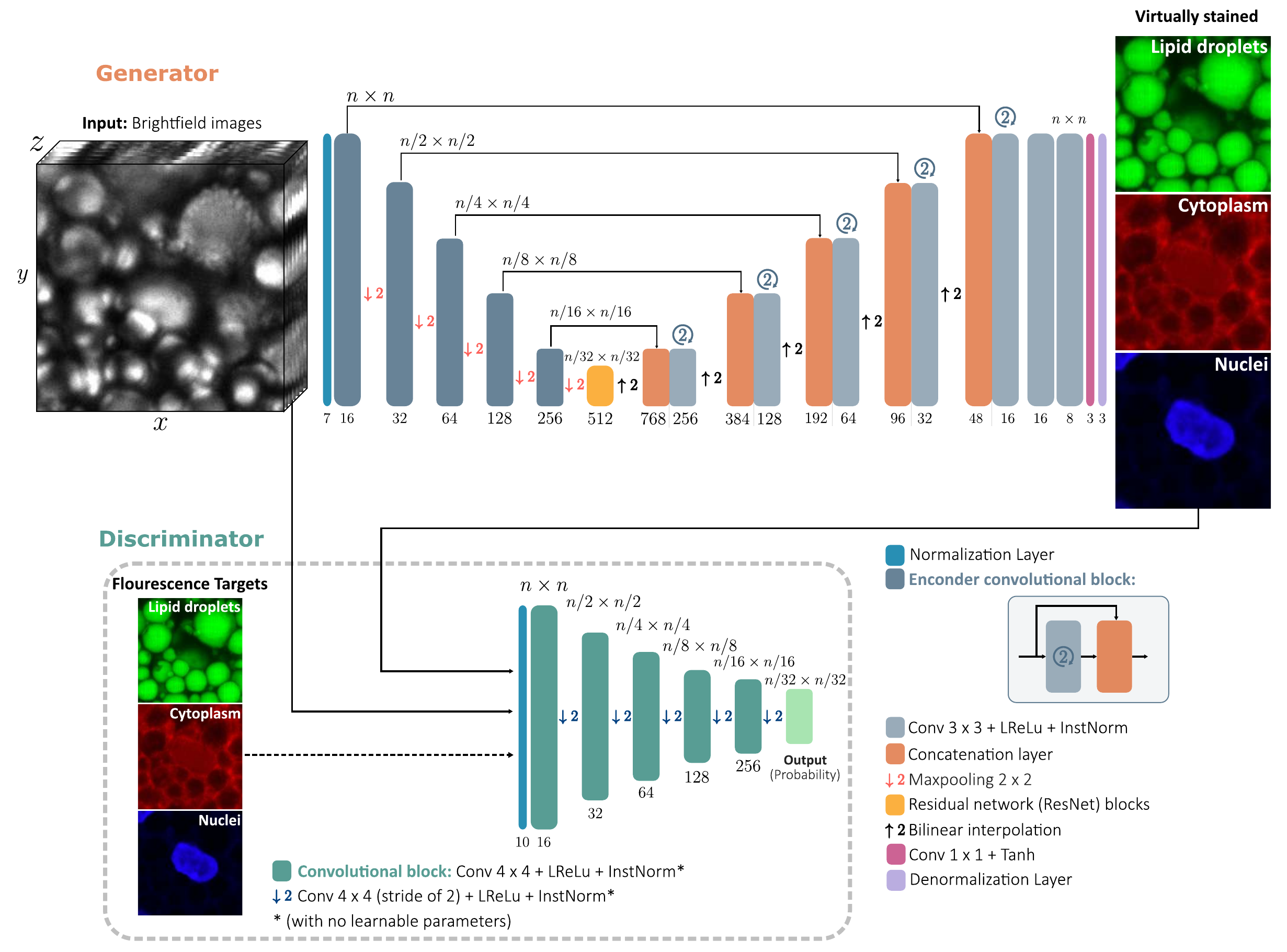}
    \caption{
    {\bf Conditional generative adversarial neural network (cGAN) for virtual staining.}
    The \emph{generator} transforms an input stack of brightfield images into virtually-stained fluorescence images of lipid droplets, cytoplasm, and nuclei, using a U-Net architecture with the most condensed layer being replaced by two residual network (ResNet) blocks \cite{he2016deep}. 
    In the first layer of the generator, we normalize each input channel (i.e., each brighfield $z$-slice) in the range $[-1,1]$ using Equation~\eqref{eq:normalization}. 
    The U-Net encoder consists of convolutional blocks followed by max-pooling layers for downsampling.
    Each convolutional block contains two paths (a sequence of two $3 \times 3$ convolutional layers, and the identity operation), which are merged by concatenation. 
    The U-Net decoder uses bilinear interpolations for upsampling followed by concatenations layers and convolutional blocks. Next, the hyperbolic tangent activation transforms the output to the range $[-1, 1]$.  
    In the last layer of the U-Net, the network learns to denormalize the output images back to original pixel values by scaling and adding an offset to the output. 
    Every layer in the generator, except the last two layers and the pooling layers, is followed by an instance normalization and a leaky ReLU activation.
    The \emph{discriminator} is designed similar to the PatchGan discriminator \cite{isola2017image} and receives both the brightfield images and fluorescence images (either the target fluorescence images or those predicted by the generator). 
    The inputs to the discriminator are normalized as those to the generator. 
    The discriminator convolutional blocks consist of $4 \times 4$ strided convolutions for downsampling. 
    In all layers in the discriminator, we use instance normalization (with no learnable parameters) and leaky ReLU activation. 
    Finally, the discriminator outputs a matrix containing the predicted probability for each patch of $32 \times 32$ pixels.
    }
    \label{fig:2}
\end{figure*}

In this work, we employ a conditional GAN (cGAN) \cite{mirza2014conditional}. 
A schematic of its architecture is shown in Figure~\ref{fig:2}. 
The generator receives as input a stack of brightfield images of the same field of view acquired at different $z$-positions and generates virtually-stained fluorescence images of lipid droplets, cytoplasm, and nuclei. 
The discriminator attempts to distinguish the generated images from fluorescently-stained samples, classifying them as either real or synthetic data. 
The \emph{conditional} aspect of the cGAN refers to the fact that the discriminator receives both the brightfield stack and the stained images as inputs. 
Thus, the task of the discriminator is \emph{conditioned} on the brightfield images, i.e., instead of answering ``is this a real staining?'', the discriminator answers ``is this a real staining \emph{for this stack of brightfield images}?''

In our implementation, the generator is based on the U-Net-architecture \cite{ronneberger2015u}, where the input image is first downsampled to a smaller representation and then upsampled to its original size, with skip connections between the downsampling and upsampling paths to retain local information.
We have modified the original U-Net architecture to optimize its performance for virtual staining. 
First, each encoder convolutional block (Figure~\ref{fig:2}) concatenates its input with the result of two sequential convolutional layers before downsampling; this helps the network to propagate information deeper, because it preserves the input information without the need for the convolutional layers to learn to preserve it.
Second, in order to tackle the vanishing gradient problem and to improve the latent space representation (i.e. a low-dimensional representation of the input data, usually corresponding to the innermost layers of the U-Net where the input is most compressed) \cite{nygate2020holographic,midtvedt2020quantitative, lei2020wavelet}, we have implemented the bottleneck of the U-Net architecture using two residual networks blocks (ResNet blocks, which preserve information from the previous layer in the network, like the encoder convolutional blocks, but add the input and output of the block, instead of concatenating them \cite{he2016deep}, Figure~\ref{fig:2}), each with 512 feature channels. 
Third, every layer (except the last two) uses instance normalization and a leaky ReLU activation (defined as $\Phi(x)=\alpha \cdot x$, where $\alpha=1$ for $x>0$ and $\alpha=0.1$ for $x<0$), which, differently from standard ReLU, has the advantage of retaining a gradient in the backpropagation step even for negative layer outputs \cite{xu2015empirical}.

In the first layer of the generator, we normalize the input brightfield $z$-stack as
\begin{equation}
    \hat{x}^i={\rm tanh}\left[
        2\frac{x^i- q^i_{1} }
            {q^i_{99} - q^i_{1}} - 1
        \right],
\label{eq:normalization}
\end{equation}
where $x^i$ is the pixel value of the $i$-th $z$-slice of the original stack and $\hat{x}^i $ is that of the rescaled $z$-slice, while $q^i_{p}$ denotes the $p$-th percentile pixel value of that $z$-slice calculated on the entire training set. 
By estimating the percentiles on the entire training set instead of on a patch-by-patch basis, the normalization becomes more resilient to outliers. 
Furthermore, normalizing by using statistical properties of the distribution of intensities rather than the minimum and maximum of the intensities, we prevent the normalization from depending on the image size and we preserve a local correspondence between the intensities of the different channels, which aids the training procedure. 
Finally, the choice of the hyperbolic tangent as a normalization function ensures that all values fall in the range $[-1, 1]$, while mitigating the effect of outliers in the intensity distribution.
In the last layer of the U-Net, the network learns to denormalize the output images back to original pixel values by scaling and offsetting the output.

We employ a discriminator that follows a conditional PatchGan architecture \cite{isola2017image}: It receives the stack of brightfield images and the fluorescence images (either the target fluorescence images or the virtually-stained images), divides them into overlapping patches, and classifies each patch as real or fake (rather than using a single descriptor for the whole input). 
This splitting arises naturally as a consequence of the discriminator’s convolutional architecture \cite{foster2019generative}. 
As shown in Figure~\ref{fig:2}, the discriminator’s convolutional blocks consist of $4 \times 4$ convolutional layers followed by strided convolutions for downsampling. 
In all layers, we use instance normalization (with no learnable parameters) and leaky ReLU activation. 
Finally, the discriminator output is a matrix that represents the predicted classification probability for each patch. 
The benefit of using a PathGAN is that the discriminator evaluates the input images based on their style rather than their content. 
This modification makes the generator task of fooling the discriminator more specialized, thus improving the quality of the generated virtual stainings \cite{nygate2020holographic}.

We have implemented this neural network using DeepTrack 2.0, an open-source software for quantitative microscopy using deep learning that we have recently developed \cite{midtvedt2020quantitative, DeepTrack},  which uses a Python-based TensorFlow backend \cite{chollet2015keras, abadi2016tensorflow}. 

\subsection*{Training procedure}

Once the network architecture is defined, we need to train it using $z$-stacks of brightfield images for which we know the corresponding target fluorescence images. 
As we have seen above, the dataset consists of 96 sets of images (each consisting of seven brightfield images and three fluorescence targets with $2560 \times 2160$ pixels). We randomly split these data into a training dataset and a validation dataset,  corresponding to 81 and 15 sets of images, respectively. 

Before starting the training process, the brightfield images and corresponding fluorescence targets need to be carefully aligned (a slight misalignment results from the different optics employed to capture the brightfield and fluorescence images). 
We use a Fourier-space correlation method that calculates a correction factor in terms of a pixel offset and a scale factor (see code in the supplementary information \cite{VirtualStainingRepo}).
Afterward, we stochastically extract $512 \times 512$ pixel patches from the corrected images and augment the training dataset using rotational and mirroring augmentations. Importantly, the misalignment must be corrected before the augmentation step because otherwise the augmentations would introduce irreducible errors and put a fundamental limit on high-frequency information.

During training, the trainable parameters of the neural network (i.e., the weights and biases of the artificial neurons in the neural network layers) are iteratively optimized using the back-propagation training algorithm \cite{mcclelland1986parallel} to minimize the loss function, i.e., the difference between virtually-stained images and target chemically-stained images. 
Initially, we set the weights of the convolutional layers of both the generator and discriminator to be randomly (normally) distributed with a mean of 0 and a standard deviation of 0.02; all of the biases are set to 0.  

In each training step, we alternately train the generator and the discriminator. 
First, the generator is tasked with predicting the fluorescence images corresponding to stacks of brightfield images. 
Then, the discriminator receives both the brightfield images and fluorescence images (either the target fluorescence images or the virtually-stained images predicted by the generator) and classifies them as real (chemically-stained images, labeled with $1$'s) or fake (virtually-stained images, labeled with $0$'s). 

The loss function of the generator is
\begin{equation}
\begin{aligned}
    \mathcal{L}_{\rm gen}  ={} & \beta  \cdot \textup{MAE}\left \{ z_{\textup{target}}, z_{\textup{output}} \right \} \\
    & +  \left (1 -  \beta \right) \cdot \left ( 1 - \textup{D}(z_{\textup{output}}) \right)^2 \enspace,
\end{aligned}
\label{eq:generator_loss}
\end{equation}
where $z_{\rm target}$ represents the chemically-stained (target) images, $z_{\rm output}$ represents the virtually-stained (generated) images, $\textup{MAE}\left \{ z_{\textup{target}}, z_{\textup{output}} \right \}$ is the mean absolute error between the target and generated images, $\textup{D}(\cdot)$ is the discriminator prediction, and $\beta$ is a weighting factor between the two part of the loss function (we set $\beta = 0.001$, which makes the typical value of the MAE roughly half the discriminator term).
Importantly, $\mathcal{L}_{\rm gen}$ depends on the discriminator prediction and penalizes the generator for producing images classified as fake.
The loss function of the discriminator is
\begin{equation}
    \mathcal{L}_{\rm disc}  = \textup{D}\left ( z_{\textup{output}} \right )^2 + \left ( 1 - \textup{D}(z_{\textup{target}}) \right)^2,
    \label{eq:discriminator_loss}
\end{equation}
which penalizes the discriminator for misclassifying real images as generated or generated images as real. Thus, the generator tries to minimize its loss by achieving $D(z_{\textup{output}}) = 1$ for the images it generates, while the discriminator tries to achieve $D(z_{\textup{output}}) = 0$ for generated images and $D(z_{\textup{target}}) = 1$ for the chemically-stained fluorescence targets. 
This leads to an adversarial behavior between the generator and the discriminator. 

We have trained both networks for 8000 epochs (each consisting of 24 batches of 8 images) using the Adam optimizer \cite{kingma2014adam} with a learning rate of $0.0002$ and $\beta_1 = 0.5$ (the exponential decay rate for the 1-st moment estimates). 
Each epoch takes 10 seconds on a NVIDIA A100 GPU (40 GB VRAM, 2430 MHz effective core clock, 6912 CUDA cores), for a total training time of about 22 hours.

\begin{figure*}[t!]
    \centering
    \includegraphics[width=1\textwidth]{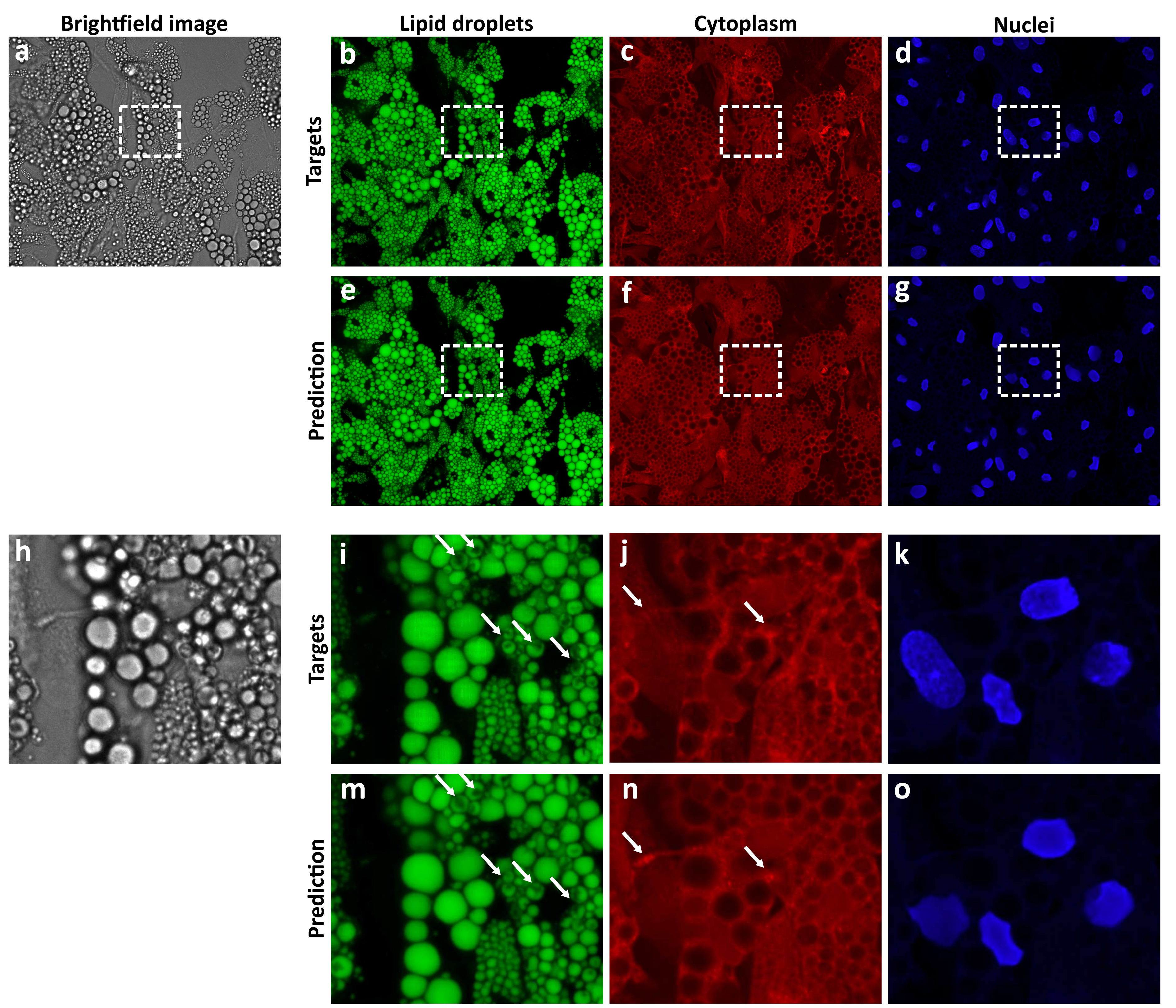}
    \caption{
    {\bf Virtually-stained fluorescence images.} 
    {\bf a} Brightfield image and corresponding {\bf b}-{\bf d} chemically-stained and {\bf e}-{\bf g} virtually-stained fluorescence images for lipid droplets, cytoplasm and nuclei.
    {\bf h}-{\bf o} Enlarged crops corresponding to the dotted boxes in {\bf a}-{\bf g}. 
    The lipid droplets are clearly visible in the brighfield image ({\bf a} and {\bf h}) thanks to their high refractive index so that the cGAN manages to accurately predict the chemically-stained images ({\bf b} and {\bf i}) generating accurate virtual stainings ({\bf e} and {\bf m}), even reproducing some details of the internal structure of the lipid droplets (indicated by the arrows in {\bf i} and {\bf m}). 
    The chemical staining of the cytoplasm ({\bf c} and {\bf j}) is also closely reproduced by the virtual staining ({\bf f} and {\bf n}).
    The virtually-stained nuclei ({\bf g} and {\bf o}) deviate more prominently from the chemically-stained ones ({\bf d} and {\bf k}), especially in the details of both their shape and texture, which can be explained by the fact that the nuclei are not clearly visible in the brightfield image so that the cGAN seems to use the surrounding cell structures to infer the presence and properties of the nuclei shape.
    }
    \label{fig:3}
\end{figure*}

\subsection*{Qualitative analysis}

Figure~\ref{fig:3} shows a representative example of virtual staining for one of the validation data realized with the cGAN described in the previous section (images for all validation data are available in the supplementary information \cite{VirtualStainingRepo}).
Figure~\ref{fig:3}a shows the first of the seven brightfield slices used as input for the cGAN, and Figure~\ref{fig:3}b, \ref{fig:3}c, and \ref{fig:3}d show the corresponding target chemically-stained fluorescence images.
Comparing the brightfield inputs with the fluorescence targets, it can be seen that the brightfield image contains information about the cellular structures, but such information is less readily accessible than in the fluorescence images.
Furthermore, it can be noticed that different cell structures have different prominence in the brightfield image, with the lipid droplets being more clearly visible than the cytoplasm, and, in turn, the cytoplasm clearer than the nuclei.

Despite the limited information in the brightfield image, the cGAN manages to predict the fluorescence targets, as can be seen in Figures~\ref{fig:3}e, \ref{fig:3}f, and \ref{fig:3}g for lipid droplets, cytoplasm, and nuclei, respectively.
Overall, the virtually-stained images appear to be qualitatively very similar to the chemically-stained ones.
Figures~\ref{fig:3}h-o show some enlarged crops of Figures~\ref{fig:3}a-g, where details can be more clearly appreciated.

The lipid droplets are virtually stained with great detail, as can be appreciated by comparing the enlarged crop of the chemical staining (Figure~\ref{fig:3}i) with that of the virtual staining (Figure~\ref{fig:3}m). 
This is to be expected because the lipid droplets, consisting primarily of lipids at high concentration, have a higher refractive index than most other intracellular objects \cite{Yanina:18}, which makes them clearly visible in the brightfield images.
Interestingly, even some details about the internal structure of the lipid droplets can be seen in the virtual staining (e.g., those indicated by the arrows in Figure~\ref{fig:3}i and \ref{fig:3}m).
These structures are probably due to proteins embedded in the surface or core of the droplets that affect the appearance of the chemically-stained cells \cite{robenek2004lipids}:
Since most of the space inside adipocytes is occupied by lipid droplets, when these cells need to increase their metabolic activity (e.g., during protein synthesis), they rearrange their contents creating textural imprints on the surfaces of the lipid droplets resulting in golf-ball-like textures. 

A lot of detail can also be found in the virtually-stained cytoplasm, as can be seen by comparing the enlarged chemically-stained image (Figure~\ref{fig:3}j) with the corresponding enlarged virtually-stained image (Figure~\ref{fig:3}n).
Similar to the lipid droplets, the high quality of the cytoplasm virtually-stained images is also to be expected as the cytoplasm also has good contrast in the brightfield images, although less than the lipid droplets. 
We can see that some of the fine structures appear to be slightly different. 
This is particularly evident in the contrast
between various cytoplasmic structures (see, e.g., those indicated by the arrows in Figure~\ref{fig:3}j and \ref{fig:3}n).  However, since the cytoplasm dye (CellTracker Deep Red) reacts with amine groups present in intracellular proteins dispersed in the cytoplasm, this propably leads to uneven staining patterns in the chemically-stained image, which are intrinsically random and, therefore, cannot be reproduced by the virtual-staining procedure. 

The nuclei are more difficult to virtually stain because they have very similar refractive index to the surrounding cytoplasm \cite{Schurmann2016}, so that there is little information about them in the brightfield image.
Nevertheless, the cGAN manages to identify them, as can be seen by comparing the enlarged crop of the chemically-stained nuclei (Figure~\ref{fig:3}k) with the corresponding virtual staining (Figure~\ref{fig:3}o), although without resolving the details of their internal structure.
The cGAN seems to extract information about the nuclei shape mostly based on the surrounding cell structures, making it difficult to predict nuclei that are not surrounded by lipid droplets.
Considering that the cell is typically at its thickest around the position of the nucleus, complementing the brightfield images with phase contrast images may give additional information that is helpful for increasing the robustness of the virtual nuclei staining.

\begin{figure*}[tbp]
    \centering
    \includegraphics[width=.84\textwidth]{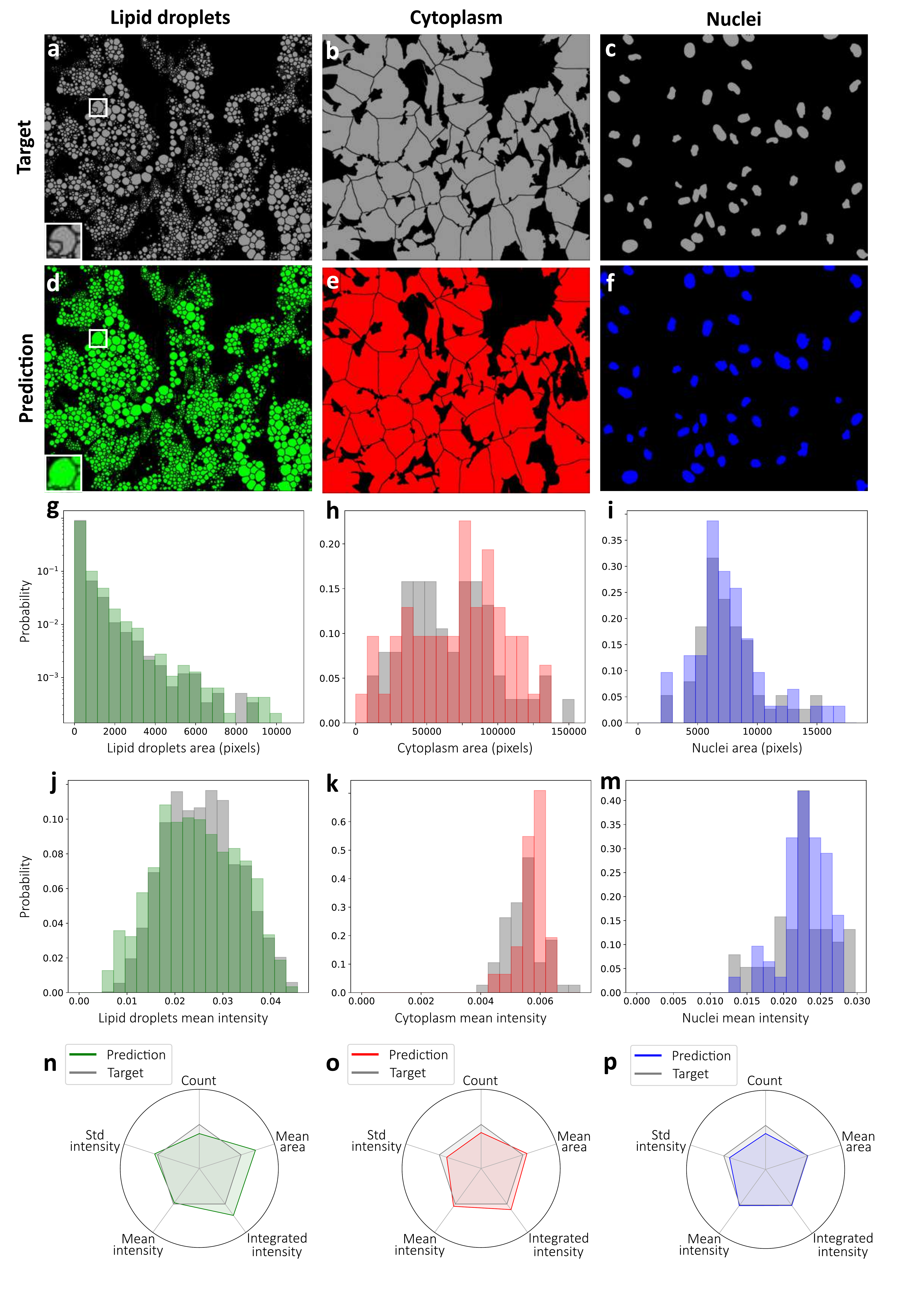}
    \caption{
    {\bf Quantitative information from chemically-stained and virtually-stained fluorescence images.}
    Segmentation obtained using CellProfiler (https://cellprofiler.org, version 4.07 \cite{mcquin2018cellprofiler}) of {\bf a}-{\bf c} chemically-stained target image and {\bf d}-{\bf f} virtually-stained generated image for lipid-droplet, cytoplasm and nuclei.
    Probability distribution for {\bf g}-{\bf i} the size and {\bf j}-{\bf m} the mean intensity of the individual lipid droplets, cytoplasmatic regions, and nuclei identified by CellProfiler for the chemically-stained (gray histograms) and virtually-stained (colored histograms) segmentations.
    {\bf n}-{\bf p} Total cell structure count, mean area, integrated intensity, mean intensity, and standard deviation of the mean intensity for the lipid droplets, cytoplasmatic regions, and nuclei identified by CellProfiler in the virtually-stained segmentations (colored outlines) normalized to those identified in the chemically-stained segmentations (gray outlines).
    }
    \label{fig:4}
\end{figure*}

\begin{table*}[t!]
    \centering
    \footnotesize
    \begin{tabular}{c|c|c|c|c|c}
        {\bf Metrics} & {\bf Target} & {\bf Prediction} & {\bf MAE} & {\bf MAE (\%)} & $\rho$ \\
        \hline 
        {\bf Lipid droplets} & & & & \\
        Pixel-value 
        & 1300$\pm$180
        %(926.2, 1707.3) 
        & 1300$\pm$170
        %(891.9, 1600.0) 
        & 60$\pm$49
        & 4.5$\pm$3.4\%
        & 0.90\\
        \rowcolor{gray!20} Count 
        & 6600$\pm$600
        %(5940, 7533) 
        & 5000$\pm$250
        %(4704, 5496) 
        & 1600$\pm$430
        %(1144, 2279) 
        & 23$\pm$7.5\% 
        & 0.82\\
        \rowcolor{gray!20}Mean area 
        & 400$\pm$42 
        %(339, 456) 
        & 570$\pm$38 
        %(499, 621)
        &
        170$\pm$30
        %(131, 222)
        & 43$\pm$15\%
        & 0.73 \\
        \rowcolor{gray!20}Integrated intensity
        & 13$\pm$2.0 
        %(10, 17) 
        & 18$\pm$2.1 
        %(14, 20) 
        & 4.9$\pm$ 1.7 
        %(2.5, 8.0) 
        & 38$\pm$16\%
        & 0.66 \\
        Mean intensity 
        & 0.025$\pm$0.0031
        %(0.021, 0.032) 
        & 0.024$\pm$0.0025
        %(0.0193, 0.0280) 
        & 0.0015$\pm$0.0012
        %(0.0002, 0.0037) 
        & 5.8$\pm$4.2\% 
        & 0.81 \\
        Std intensity 
        & 0.0035$\pm$0.00037
        %(0.0029, 0.0041) 
        & 0.0038$\pm$0.00022 
        %(0.0035, 0.0042) 
        & 0.0003$\pm$0.00024
        %(0.00001, 0.0008) 
        & 11$\pm$8.0\%
        & 0.76 \\
        \hline 
        {\bf Cytoplasm} & & & & \\
        Pixel-value 
        & 320$\pm$17
        %(305.0, 341.8) 
        & 330$\pm$13
        %(322.1, 345.0) 
        & 13$\pm$10
        & 4.1$\pm$2.9\%
        & 0.50 \\
        \rowcolor{gray!20} Count 
        & 34$\pm$6.2
        %(25, 44) 
        & 31$\pm$4.0
        %(26, 38) 
        & 3.9$\pm$3.0
        %(0, 8) 
        &11$\pm$7.2\% 
        & 0.74 \\
        \rowcolor{gray!20} Mean area 
        & 79000$\pm$11000
        %(66229, 103364) 
        & 84000$\pm$11000
        %(70992, 104004) 
        & 6800$\pm$4400
        %(112, 12983) 
        & 8.8$\pm$5.9\%
        & 0.82 \\
        Integrated intensity 
        & 430$\pm$59
        %(377, 563) 
        & 470$\pm$59
        %(401, 569) 
        & 45$\pm$28
        %(2, 85)
        & 10$\pm$6.5\% 
        & 0.77 \\
        Mean intensity 
        & 0.0055$\pm$0.00016
        %(0.0053, 0.0058) 
        & 0.0056$\pm$0.00017
        %(0.0054, 0.0059) 
        & 0.00012$\pm$0.00012
        %(0.00002, 0.00036)
        & 2.3$\pm$2.0\%
        & 0.55 \\
        Std intensity
        & 0.0016$\pm$0.000086
        %(0.0015, 0.0017) 
        & 0.0013$\pm$0.000064
        %(0.0012, 0.0014) 
        & 0.00033$\pm$0.000077
        %(0.0002, 0.0005)
        & 20$\pm$6.5\% 
        & 0.55 \\
        \hline 
        {\bf Nuclei} & & & & \\
        Pixel-value 
        & 290$\pm$9.0
        %(272.1, 327.8) 
        & 280$\pm$7.0
        %(254.5, 298.6) 
        & 17$\pm$6.0
        & 5.6$\pm$1.9\%
        & 0.65 \\
        \rowcolor{gray!20} Count 
        & 34$\pm$6.2
        %(25, 44) 
        & 31$\pm$4.0
        %(26, 38) 
        & 3.9$\pm$3.0
        %(0, 8)
        & 11$\pm$7.2\% 
        & 0.74 \\
        \rowcolor{gray!20} Mean area 
        & 7400$\pm$1200
        %(6227, 10116) 
        & 7100$\pm$1200
        %(5438, 9490) 
        & 430$\pm$300
        %(28, 898) 
        & 5.9$\pm$4.3\%
        & 0.92 \\
        Integrated intensity 
        & 170$\pm$27 
        %(140, 231) 
        & 160$\pm$25
        %(124, 202) 
        & 14$\pm$10
        %(2, 30)
        & 8.3$\pm$6.1\%
        & 0.83 \\
        Mean intensity 
        & 0.022$\pm$0.00092
        %(0.0201, 0.0232) 
        & 0.022$\pm$0.00098
        %(0.0203, 0.0235) 
        & 0.00082$\pm$0.00071
        %(0.0001, 0.0022)
        & 3.7$\pm$3.0\%
        & 0.35 \\
        Std intensity 
        & 0.0065$\pm$0.00049
        %(0.0054, 0.0070) 
        & 0.0053$\pm$0.00049 
        %(0.0044, 0.0060) 
        & 0.0012$\pm$0.00053
        %(0.0004, 0.0021) 
        & 18$\pm$7.7\%
        & 0.38 \\
    \end{tabular}
    \caption{
    {\bf Comparison of features extracted from chemically-stained and virtually-stained images for the whole validation dataset.}
    Average and standard deviation of various metrics (pixel value, count, mean area, integrated intensity, mean intensity, and standard deviation of the mean intensity of lipid droplets, cytoplasmic regions, and nuclei) calculated over the 15 sets of target chemically-stained images and of the predicted virtually-stained images of the validation dataset.
    We also report the value and percentage of the mean absolute error (MAE) as well as the Pearson correlation $\rho$ between the metrics calculated on the target and predicted images.
    Note that the pixel values are in the original image range $[0, 65535]$, while the intensity measurements are extracted with CellProfiler from images rescaled in the range $[0, 1]$. The features that are most biologically relevant for each cell structure are highlighted in gray.
    }
    \label{tab:1}
\end{table*}

\subsection*{Quantitative analysis}

The stained images are then used to extract quantitative biological information about the cell structures. 
For example, quantitative information about the cellular lipid droplet content is critical to study metabolic diseases where the fat storage in adipocytes plays a pivotal role and to dissect the mechanisms leading to organ injury due to lipid deposition in ectopic tissue \cite{van2008lipid}. 
As a consequence, generation of accurate and relevant quantitative cell structure data is of key importance for biomedical and pharmaceutical research as well as for clinical therapeutic decisions.

Here, we have used the open-source software CellProfiler (version 4.07 \cite{mcquin2018cellprofiler}) to identify and segment the lipid droplets, cytoplasm and nuclei in both the chemically-stained and virtually-stained fluorescence images (the analysis pipeline is available in the supplementary information \cite{VirtualStainingRepo}). 
For each cell structure, we employ a feature-extraction pipeline that calculates the number of cell structures in each image, their mean area in pixels, their integrated intensity, their mean intensity, and the standard deviation of their mean intensity. 
The results of this quantitative analysis are shown in Figure~\ref{fig:4} for the same representative set of validation images used in Figure~\ref{fig:3} (the results for all validation data are available in the supplementary information \cite{VirtualStainingRepo}).
The aggregated results for the whole validation dataset are presented in Table~\ref{tab:1}.

The first step of the feature-extraction pipeline is to segment the relevant cell structures. 
Starting from the fluorescence images, the feature-extraction pipeline identifies relevant cellular structures based on threshold values for intensity, size and shape. 
Figures~\ref{fig:4}a-c show the segmentations obtained from the chemically-stained images, and Figures~\ref{fig:4}d-e the corresponding segmentations obtained from the virtually-stained images.

In the feature-extraction pipeline, the nuclei are identified first (Figures~\ref{fig:4}c and ~\ref{fig:4}f). 
Since the lipid droplets in the adipocytes may occlude the nuclei and physically change their size and shape, a wide range of possible nuclear diameters and shapes is selected to ensure a successful segmentation. 
Furthermore, since the intensity of the nuclei varies, an adaptive thresholding strategy is chosen (i.e., for each pixel, the threshold is calculated based on the surrounding pixels within a given neighborhood). 
As a last step, nuclei that are clumped together are distinguished by their shape.
Identifying the nuclei is critically important because the number of nuclei is often used for the quantification of different biological phenomena, for example the average amount of lipids per cell in the context of diabetes research. 

In the second part of the feature-extraction pipeline, the cytoplasm is segmented to determine the cell boundaries, starting from the locations of the previously identified nuclei (Figures~\ref{fig:4}b and ~\ref{fig:4}e). 
An adaptive thresholding strategy is again used, with a larger adaptive window (the neighborhood considered for the calculation of the threshold) compared to that used for the nuclei segmentation. Identifying the cytoplasm structure is important because it gives information about the cell size (measured area) and morphology (e.g., presence of protrusions or blebbing features), which are in turn related to the physiological state of the cell \cite{charras2008short}.

In the final part of the feature-extraction pipeline, the lipid droplets are segmented independently from the nuclei and cytoplasm (Figures~\ref{fig:4}a and ~\ref{fig:4}d).
This segmentation is done in two steps to target separately the smaller and larger lipid droplets. 
For each of the two steps, a range of expected diameters and intensities are selected for the image thresholding. 
Since lipid droplets in each of the size distributions have similar peak intensities, a global thresholding strategy is used for their identification. 
Lipid droplets that are clumped together are distinguished by their intensity rather than their shape, which is consistently round for all the lipid droplets. 

The segmented images are then used to count and characterize the cell structures.
Figures~\ref{fig:4}g-m show that there is a good agreement between the probability distribution histograms for the area size and mean intensity of the cell structures identified from the chemically-stained (gray histograms) and virtually-stained (colored histograms) segmentations for lipid droplets (Figures~\ref{fig:4}g and \ref{fig:4}j), cytoplasm (Figures~\ref{fig:4}h and \ref{fig:4}k), and nuclei (Figures~\ref{fig:4}i and \ref{fig:4}m).
Figures~\ref{fig:4}n-p show the cell structure count in the image, their mean area, their combined integrated intensity over the image, the mean intensity of cell structures in the image, and the standard deviation of  the  mean  intensity identified by CellProfiler in the virtually-stained images (colored outlines) normalized to those identified in the chemically-stained images (gray outlines) for the lipid droplets (Figure~\ref{fig:4}n), cytoplasmic regions (Figure~\ref{fig:4}o), and nuclei (Figure~\ref{fig:4}p).

The aggregated results for the features extracted using CellProfiler for the whole validation dataset are presented in Table~\ref{tab:1}.
Importantly, there is a high correlation (Pearson correlation coefficient $\rho$ in Table~\ref{tab:1}) between all metrics obtained with the chemically-stained and virtually-stained images.
This indicates that any deviation between these metrics is systematic and consistent, which is highly relevant for biological experiments, where the focus is often not on absolute values but rather on the comparison of different samples.

The feature extraction from the virtually-stained images shows the best performance for the lipid droplets.
This is very useful for potential applications because, e.g., lipid droplets are often used to measure the effect of drugs for metabolic diseases. In this context, the amount of fat in cells can be quantified by normalizing the number of lipid droplets, their mean area or integrated intensity to the number of cells in the image.
A systematically lower number of larger lipid droplets is identified in the segmented virtually-stained images (Figures~\ref{fig:4}d) compared to the segmented chemically-stained images (Figures~\ref{fig:4}a).
This can be partly explained by the fact that chemically-stained fluorescence images of the lipid droplets have some intensity variations (see, e.g., those indicated by the arrows in Figures~\ref{fig:3}i and ~\ref{fig:3}m), which may result in the erroneous segmentation of a single lipid droplet into multiple parts (see, e.g., the inset in Figure~\ref{fig:4}a).
Even though these intensity variations are reproduced in the virtually-stained images (see, e.g., those indicated by the arrows in Figures~\ref{fig:3}j and ~\ref{fig:3}n), they do not translate into an erroneous segmentation of the image by CellProfiler,
leading to identification of fewer but larger lipid droplets (see, e.g., the inset in Figure~\ref{fig:4}d). Therefore, the lipid droplet count is lower, their area larger, and their integrate intensity is higher when analyzing the virtually-stained images compared to when analyzing the chemically-stained ones (Table~\ref{tab:1}).
Nevertheless, the average and standard deviation of their mean intensity are more closely estimated (probably thanks to the fact that these are intensive quantities).

The main information extracted from the cytoplasm staining is related to the cell boundaries and morphology.
In this respect, the cell count and mean area are the most important metrics, which are reproduced very well by the analysis of the virtually-stained images (Table~\ref{tab:1}).
The other metrics are related to the intensity of the cytoplasm, which can be inconsistent even in the chemically-stained images because the cytoplasmic dye (CellTracker Deep Red) reacts with amine groups present in intracellular proteins dispersed in the cytoplasm producing an uneven texture. 
This explains why the cGAN cannot predict the exact spatial distribution and amount of the chemical dye from which the chemically-stained images are obtained.
On the other hand, the metrics about the integrated intensity, mean intensity, and  standard deviation of the mean intensity are reproduced accurately from the virtually-stained images.

The nuclei are used to identify the individual cells, for which both the number and morphological properties of the nuclei are needed.
In this respect, the most important measures are the nuclei count and mean area, which are determined accurately using the virtually-stained images (Table~\ref{tab:1}).
The other metrics (pixel value, mean intensity, and standard deviation of the intensity) are less comparable to the chemically-stained fluorescence images. 
The cGAN does not manage entirely to capture the dynamic content of the nuclei, possibly because of the non-static chromatin conformations present in living cells, resulting in different levels of dye accessibility. 
With this information not being visible in the brightfield images, it is not surprising that the virtual staining does not include textural details.
Nevertheless, this is not generally a problem because in most studies the cell nuclei morphology or chromatin conformation is not the aim, rather the nuclei constitute cell structures useful as normalization factors. The virtual staining does offer sensitive cell number determination and, as such, enables cell-cell comparison of other measured parameters. 
Considering the known phototoxicity of Hoechst 33342 in time-lapse imaging series of living cells \cite{purschke2010phototoxicity}, and the cGAN enables accurate nuclear counts and cell segmentation, and may be preferred over chemical staining.

\section*{Conclusions}

We have developed a deep-learning-powered method for quantitative analysis of intracellular structures in terms of their size, morphology, and content. 
The method is based on virtually-stained images of cells derived from brightfield images and subsequent downstream analysis to quantify the properties of the virtually-stained cell structures. 

We have demonstrated the accuracy and reliability of our method by virtually staining and quantifying the lipid droplets, cytoplasm, and cell nuclei from brightfield images of stem-cell-derived adipocytes. While the lipid droplets are easily visible in the brightfield images, direct quantification of their size and content using conventional analysis techniques is challenging, and fluorescent staining techniques are typically used. The cytoplasm and cell nuclei are almost indistinguishable based on their optical contrast, but also in this case the neural network manages to reconstruct them, probably also making use of information contained in the spatial distribution of the lipid droplet.

Compared to standard approaches based on fluorescent staining, our approach is less labor-intensive and the results do not depend on careful optimization of the staining procedure. 
Therefore, the results are more robust and can potentially be compared across experiments and even across labs.
We note also that the proposed approach is not limited to the structures quantified in this work, but can be applied to virtually stain and quantify any intracellular object with unique optical characteristics.
Furthermore, virtual staining does not exclude fluorescent imaging, so additional information can also be obtained from the liberated fluorescence channels, such as particle uptake or protein expression, both of which are important, e.g., for subcutaneous dosing of nanomedicines and vaccines.

To make this method readily available for future applications, we provide a Python open-source software package, which can be  personalized and optimized for the needs of specific users and applications \cite{VirtualStainingRepo}.

\section*{Funding and acknowledgments}

The authors would like to thank Anders Broo and Lars Tornberg from AstraZeneca and Johanna Bergman and Sheetal Reddy from AI Sweden for enlightening discussions. 
AI Sweden provided access to their computational resources. 
The authors would also like to acknowledge that the idea for this work is inspired by the \textit{Adipocyte Cell Imaging Challenge} held by AI Sweden and AstraZeneca. 
This work was partly supported by the H2020 European Research Council (ERC) Starting Grant ComplexSwimmers (677511), the Knut and Alice Wallenberg Foundation, and the Swedish Strategic Research Foundation (ITM17-0384).

\section*{Author Contributions}
Author contributions are defined based on the CRediT (Contributor Roles Taxonomy) and listed alphabetically. 
Conceptualization: C.B.A., D.M., S.R., A.S., G.V.
Formal analysis: S.H., B.M., D.M., J.P., G.V.
Funding acquisition:  C.B.A., D.M., S.R., G.V.
Investigation: S.H., B.M., D.M., J.P., A.S., G.V.
Methodology: S.H., B.M., D.M., J.P., G.V.
Project administration: G.V. 
Software: S.H., D.M., B.M., J.P.
Supervision: G.V. 
Validation: S.H., B.M., D.M., J.P., G.V.
Visualization: S.H., B.M., J.P.
Writing—original draft: S.H., B.M., D.M., J.P., G.V.
Writing—review and editing: C.B.A., S.H., B.M., D.M., J.P., S.R., A.S., G.V.

\section*{Competing Interests statement}
The authors declare no competing interests.

%\bibliography{sample}

\begin{thebibliography}{45}%
\makeatletter
\providecommand \@ifxundefined [1]{%
 \@ifx{#1\undefined}
}%
\providecommand \@ifnum [1]{%
 \ifnum #1\expandafter \@firstoftwo
 \else \expandafter \@secondoftwo
 \fi
}%
\providecommand \@ifx [1]{%
 \ifx #1\expandafter \@firstoftwo
 \else \expandafter \@secondoftwo
 \fi
}%
\providecommand \natexlab [1]{#1}%
\providecommand \enquote  [1]{``#1''}%
\providecommand \bibnamefont  [1]{#1}%
\providecommand \bibfnamefont [1]{#1}%
\providecommand \citenamefont [1]{#1}%
\providecommand \href@noop [0]{\@secondoftwo}%
\providecommand \href [0]{\begingroup \@sanitize@url \@href}%
\providecommand \@href[1]{\@@startlink{#1}\@@href}%
\providecommand \@@href[1]{\endgroup#1\@@endlink}%
\providecommand \@sanitize@url [0]{\catcode `\\12\catcode `\$12\catcode
  `\&12\catcode `\#12\catcode `\^12\catcode `\_12\catcode `\%12\relax}%
\providecommand \@@startlink[1]{}%
\providecommand \@@endlink[0]{}%
\providecommand \url  [0]{\begingroup\@sanitize@url \@url }%
\providecommand \@url [1]{\endgroup\@href {#1}{\urlprefix }}%
\providecommand \urlprefix  [0]{URL }%
\providecommand \Eprint [0]{\href }%
\providecommand \doibase [0]{http://dx.doi.org/}%
\providecommand \selectlanguage [0]{\@gobble}%
\providecommand \bibinfo  [0]{\@secondoftwo}%
\providecommand \bibfield  [0]{\@secondoftwo}%
\providecommand \translation [1]{[#1]}%
\providecommand \BibitemOpen [0]{}%
\providecommand \bibitemStop [0]{}%
\providecommand \bibitemNoStop [0]{.\EOS\space}%
\providecommand \EOS [0]{\spacefactor3000\relax}%
\providecommand \BibitemShut  [1]{\csname bibitem#1\endcsname}%
\let\auto@bib@innerbib\@empty
%</preamble>
\bibitem [{\citenamefont {Rizzuto}\ \emph {et~al.}(1995)\citenamefont
  {Rizzuto}, \citenamefont {Brini}, \citenamefont {Pizzo}, \citenamefont
  {Murgia},\ and\ \citenamefont {Pozzan}}]{rizzuto1995chimeric}%
  \BibitemOpen
  \bibfield  {author} {\bibinfo {author} {\bibfnamefont {R}~\bibnamefont
  {Rizzuto}}, \bibinfo {author} {\bibfnamefont {M}~\bibnamefont {Brini}},
  \bibinfo {author} {\bibfnamefont {P}~\bibnamefont {Pizzo}}, \bibinfo {author}
  {\bibfnamefont {M}~\bibnamefont {Murgia}}, \ and\ \bibinfo {author}
  {\bibfnamefont {T}~\bibnamefont {Pozzan}},\ }\bibfield  {title} {\enquote
  {\bibinfo {title} {Chimeric green fluorescent protein as a tool for
  visualizing subcellular organelles in living cells},}\ }\href@noop {}
  {\bibfield  {journal} {\bibinfo  {journal} {Curr. Biol.}\ }\textbf {\bibinfo
  {volume} {5}},\ \bibinfo {pages} {635--642} (\bibinfo {year}
  {1995})}\BibitemShut {NoStop}%
\bibitem [{\citenamefont {Bj{\"o}rklund}\ \emph {et~al.}(2006)\citenamefont
  {Bj{\"o}rklund}, \citenamefont {Taipale}, \citenamefont {Varjosalo},
  \citenamefont {Saharinen}, \citenamefont {Lahdenper{\"a}},\ and\
  \citenamefont {Taipale}}]{bjorklund2006identification}%
  \BibitemOpen
  \bibfield  {author} {\bibinfo {author} {\bibfnamefont {M}~\bibnamefont
  {Bj{\"o}rklund}}, \bibinfo {author} {\bibfnamefont {M}~\bibnamefont
  {Taipale}}, \bibinfo {author} {\bibfnamefont {M}~\bibnamefont {Varjosalo}},
  \bibinfo {author} {\bibfnamefont {J}~\bibnamefont {Saharinen}}, \bibinfo
  {author} {\bibfnamefont {J}~\bibnamefont {Lahdenper{\"a}}}, \ and\ \bibinfo
  {author} {\bibfnamefont {J}~\bibnamefont {Taipale}},\ }\bibfield  {title}
  {\enquote {\bibinfo {title} {Identification of pathways regulating cell size
  and cell-cycle progression by {RNA}i},}\ }\href@noop {} {\bibfield  {journal}
  {\bibinfo  {journal} {Nature}\ }\textbf {\bibinfo {volume} {439}},\ \bibinfo
  {pages} {1009--1013} (\bibinfo {year} {2006})}\BibitemShut {NoStop}%
\bibitem [{\citenamefont {Kepp}\ \emph {et~al.}(2011)\citenamefont {Kepp},
  \citenamefont {Galluzzi}, \citenamefont {Lipinski}, \citenamefont {Yuan},\
  and\ \citenamefont {Kroemer}}]{kepp2011cell}%
  \BibitemOpen
  \bibfield  {author} {\bibinfo {author} {\bibfnamefont {O}~\bibnamefont
  {Kepp}}, \bibinfo {author} {\bibfnamefont {L}~\bibnamefont {Galluzzi}},
  \bibinfo {author} {\bibfnamefont {M}~\bibnamefont {Lipinski}}, \bibinfo
  {author} {\bibfnamefont {J}~\bibnamefont {Yuan}}, \ and\ \bibinfo {author}
  {\bibfnamefont {G}~\bibnamefont {Kroemer}},\ }\bibfield  {title} {\enquote
  {\bibinfo {title} {Cell death assays for drug discovery},}\ }\href@noop {}
  {\bibfield  {journal} {\bibinfo  {journal} {Nat. Rev. Drug Discov.}\ }\textbf
  {\bibinfo {volume} {10}},\ \bibinfo {pages} {221--237} (\bibinfo {year}
  {2011})}\BibitemShut {NoStop}%
\bibitem [{\citenamefont {Lulevich}\ \emph {et~al.}(2009)\citenamefont
  {Lulevich}, \citenamefont {Shih}, \citenamefont {Lo},\ and\ \citenamefont
  {Liu}}]{lulevich2009cell}%
  \BibitemOpen
  \bibfield  {author} {\bibinfo {author} {\bibfnamefont {V}~\bibnamefont
  {Lulevich}}, \bibinfo {author} {\bibfnamefont {Y~P}\ \bibnamefont {Shih}},
  \bibinfo {author} {\bibfnamefont {S~H}\ \bibnamefont {Lo}}, \ and\ \bibinfo
  {author} {\bibfnamefont {G~Y}\ \bibnamefont {Liu}},\ }\bibfield  {title}
  {\enquote {\bibinfo {title} {Cell tracing dyes significantly change single
  cell mechanics},}\ }\href@noop {} {\bibfield  {journal} {\bibinfo  {journal}
  {J. Phys. Chem. B}\ }\textbf {\bibinfo {volume} {113}},\ \bibinfo {pages}
  {6511--6519} (\bibinfo {year} {2009})}\BibitemShut {NoStop}%
\bibitem [{\citenamefont {Alford}\ \emph {et~al.}(2009)\citenamefont {Alford},
  \citenamefont {Simpson}, \citenamefont {Duberman}, \citenamefont {Hill},
  \citenamefont {Ogawa}, \citenamefont {Regino}, \citenamefont {Kobayashi},\
  and\ \citenamefont {Choyke}}]{alford2009toxicity}%
  \BibitemOpen
  \bibfield  {author} {\bibinfo {author} {\bibfnamefont {R}~\bibnamefont
  {Alford}}, \bibinfo {author} {\bibfnamefont {H~M}\ \bibnamefont {Simpson}},
  \bibinfo {author} {\bibfnamefont {J}~\bibnamefont {Duberman}}, \bibinfo
  {author} {\bibfnamefont {G~C}\ \bibnamefont {Hill}}, \bibinfo {author}
  {\bibfnamefont {M}~\bibnamefont {Ogawa}}, \bibinfo {author} {\bibfnamefont
  {C}~\bibnamefont {Regino}}, \bibinfo {author} {\bibfnamefont {H}~\bibnamefont
  {Kobayashi}}, \ and\ \bibinfo {author} {\bibfnamefont {P~L}\ \bibnamefont
  {Choyke}},\ }\bibfield  {title} {\enquote {\bibinfo {title} {Toxicity of
  organic fluorophores used in molecular imaging: literature review},}\
  }\href@noop {} {\bibfield  {journal} {\bibinfo  {journal} {Mol. Imaging}\
  }\textbf {\bibinfo {volume} {8}},\ \bibinfo {pages} {7290--2009} (\bibinfo
  {year} {2009})}\BibitemShut {NoStop}%
\bibitem [{\citenamefont {Ounkomol}\ \emph {et~al.}(2018)\citenamefont
  {Ounkomol}, \citenamefont {Seshamani}, \citenamefont {Maleckar},
  \citenamefont {Collman},\ and\ \citenamefont {Johnson}}]{ounkomol2018label}%
  \BibitemOpen
  \bibfield  {author} {\bibinfo {author} {\bibfnamefont {C}~\bibnamefont
  {Ounkomol}}, \bibinfo {author} {\bibfnamefont {S}~\bibnamefont {Seshamani}},
  \bibinfo {author} {\bibfnamefont {M~M}\ \bibnamefont {Maleckar}}, \bibinfo
  {author} {\bibfnamefont {F}~\bibnamefont {Collman}}, \ and\ \bibinfo {author}
  {\bibfnamefont {G~R}\ \bibnamefont {Johnson}},\ }\bibfield  {title} {\enquote
  {\bibinfo {title} {Label-free prediction of three-dimensional fluorescence
  images from transmitted-light microscopy},}\ }\href@noop {} {\bibfield
  {journal} {\bibinfo  {journal} {Nat. Methods}\ }\textbf {\bibinfo {volume}
  {15}},\ \bibinfo {pages} {917--920} (\bibinfo {year} {2018})}\BibitemShut
  {NoStop}%
\bibitem [{\citenamefont {Zimmermann}(2005)}]{zimmermann2005spectral}%
  \BibitemOpen
  \bibfield  {author} {\bibinfo {author} {\bibfnamefont {T}~\bibnamefont
  {Zimmermann}},\ }\bibfield  {title} {\enquote {\bibinfo {title} {Spectral
  imaging and linear unmixing in light microscopy},}\ }in\ \href@noop {} {\emph
  {\bibinfo {booktitle} {Microscopy techniques}}}\ (\bibinfo  {publisher}
  {Springer},\ \bibinfo {year} {2005})\ pp.\ \bibinfo {pages}
  {245--265}\BibitemShut {NoStop}%
\bibitem [{\citenamefont {Lecun}\ \emph {et~al.}(2015)\citenamefont {Lecun},
  \citenamefont {Bengio},\ and\ \citenamefont
  {Hinton}}]{Lecun2015DeepLearning}%
  \BibitemOpen
  \bibfield  {author} {\bibinfo {author} {\bibfnamefont {Y}~\bibnamefont
  {Lecun}}, \bibinfo {author} {\bibfnamefont {Y}~\bibnamefont {Bengio}}, \ and\
  \bibinfo {author} {\bibfnamefont {G}~\bibnamefont {Hinton}},\ }\bibfield
  {title} {\enquote {\bibinfo {title} {{Deep learning}},}\ }\href {\doibase
  10.1038/nature14539} {\bibfield  {journal} {\bibinfo  {journal} {Nature}\
  }\textbf {\bibinfo {volume} {521}},\ \bibinfo {pages} {436--444} (\bibinfo
  {year} {2015})}\BibitemShut {NoStop}%
\bibitem [{\citenamefont {Cihos}\ \emph {et~al.}(2020)\citenamefont {Cihos},
  \citenamefont {Gustavsson}, \citenamefont {Mehlig},\ and\ \citenamefont
  {Volpe}}]{cichos2020machine}%
  \BibitemOpen
  \bibfield  {author} {\bibinfo {author} {\bibfnamefont {F}~\bibnamefont
  {Cihos}}, \bibinfo {author} {\bibfnamefont {K}~\bibnamefont {Gustavsson}},
  \bibinfo {author} {\bibfnamefont {B}~\bibnamefont {Mehlig}}, \ and\ \bibinfo
  {author} {\bibfnamefont {G}~\bibnamefont {Volpe}},\ }\bibfield  {title}
  {\enquote {\bibinfo {title} {Machine learning for active matter},}\
  }\href@noop {} {\bibfield  {journal} {\bibinfo  {journal} {Nat. Mach.
  Intell.}\ }\textbf {\bibinfo {volume} {2}},\ \bibinfo {pages} {94--103}
  (\bibinfo {year} {2020})}\BibitemShut {NoStop}%
\bibitem [{\citenamefont {Barbastathis}\ \emph {et~al.}(2019)\citenamefont
  {Barbastathis}, \citenamefont {Ozcan},\ and\ \citenamefont
  {Situ}}]{barbastathis2019use}%
  \BibitemOpen
  \bibfield  {author} {\bibinfo {author} {\bibfnamefont {G}~\bibnamefont
  {Barbastathis}}, \bibinfo {author} {\bibfnamefont {A}~\bibnamefont {Ozcan}},
  \ and\ \bibinfo {author} {\bibfnamefont {G}~\bibnamefont {Situ}},\ }\bibfield
   {title} {\enquote {\bibinfo {title} {On the use of deep learning for
  computational imaging},}\ }\href@noop {} {\bibfield  {journal} {\bibinfo
  {journal} {Optica}\ }\textbf {\bibinfo {volume} {6}},\ \bibinfo {pages}
  {921--943} (\bibinfo {year} {2019})}\BibitemShut {NoStop}%
\bibitem [{\citenamefont {Hannel}\ \emph {et~al.}(2018)\citenamefont {Hannel},
  \citenamefont {Abdulali}, \citenamefont {O’Brien},\ and\ \citenamefont
  {Grier}}]{Hannel2018Machine-learningParticles}%
  \BibitemOpen
  \bibfield  {author} {\bibinfo {author} {\bibfnamefont {M~D}\ \bibnamefont
  {Hannel}}, \bibinfo {author} {\bibfnamefont {A}~\bibnamefont {Abdulali}},
  \bibinfo {author} {\bibfnamefont {M}~\bibnamefont {O’Brien}}, \ and\
  \bibinfo {author} {\bibfnamefont {D~G}\ \bibnamefont {Grier}},\ }\bibfield
  {title} {\enquote {\bibinfo {title} {{Machine-learning techniques for fast
  and accurate feature localization in holograms of colloidal particles}},}\
  }\href {\doibase 10.1364/oe.26.015221} {\bibfield  {journal} {\bibinfo
  {journal} {Opt. Express}\ }\textbf {\bibinfo {volume} {26}},\ \bibinfo
  {pages} {15221} (\bibinfo {year} {2018})}\BibitemShut {NoStop}%
\bibitem [{\citenamefont {Newby}\ \emph {et~al.}(2018)\citenamefont {Newby},
  \citenamefont {Schaefer}, \citenamefont {Lee}, \citenamefont {Forest},\ and\
  \citenamefont {Lai}}]{Newby2018Convolutional3D}%
  \BibitemOpen
  \bibfield  {author} {\bibinfo {author} {\bibfnamefont {J~M}\ \bibnamefont
  {Newby}}, \bibinfo {author} {\bibfnamefont {A~M}\ \bibnamefont {Schaefer}},
  \bibinfo {author} {\bibfnamefont {P~T}\ \bibnamefont {Lee}}, \bibinfo
  {author} {\bibfnamefont {M~G}\ \bibnamefont {Forest}}, \ and\ \bibinfo
  {author} {\bibfnamefont {S~K}\ \bibnamefont {Lai}},\ }\bibfield  {title}
  {\enquote {\bibinfo {title} {{Convolutional neural networks automate
  detection for tracking of submicron-scale particles in 2D and 3D}},}\ }\href
  {\doibase 10.1073/pnas.1804420115} {\bibfield  {journal} {\bibinfo  {journal}
  {PNAS}\ }\textbf {\bibinfo {volume} {115}},\ \bibinfo {pages} {9026--9031}
  (\bibinfo {year} {2018})}\BibitemShut {NoStop}%
\bibitem [{\citenamefont {Helgadottir}\ \emph {et~al.}(2019)\citenamefont
  {Helgadottir}, \citenamefont {Argun},\ and\ \citenamefont
  {Volpe}}]{Helgadottir2019DigitalLearning}%
  \BibitemOpen
  \bibfield  {author} {\bibinfo {author} {\bibfnamefont {S}~\bibnamefont
  {Helgadottir}}, \bibinfo {author} {\bibfnamefont {A}~\bibnamefont {Argun}}, \
  and\ \bibinfo {author} {\bibfnamefont {G}~\bibnamefont {Volpe}},\ }\bibfield
  {title} {\enquote {\bibinfo {title} {{Digital video microscopy enhanced by
  deep learning}},}\ }\href {\doibase 10.1364/OPTICA.6.000506} {\bibfield
  {journal} {\bibinfo  {journal} {Optica}\ }\textbf {\bibinfo {volume} {6}},\
  \bibinfo {pages} {506--513} (\bibinfo {year} {2019})}\BibitemShut {NoStop}%
\bibitem [{\citenamefont {Midtvedt}\ \emph
  {et~al.}(2020{\natexlab{a}})\citenamefont {Midtvedt}, \citenamefont
  {Helgadottir}, \citenamefont {Argun}, \citenamefont {Pineda}, \citenamefont
  {Midtvedt},\ and\ \citenamefont {Volpe}}]{midtvedt2020quantitative}%
  \BibitemOpen
  \bibfield  {author} {\bibinfo {author} {\bibfnamefont {B}~\bibnamefont
  {Midtvedt}}, \bibinfo {author} {\bibfnamefont {S}~\bibnamefont
  {Helgadottir}}, \bibinfo {author} {\bibfnamefont {A}~\bibnamefont {Argun}},
  \bibinfo {author} {\bibfnamefont {J}~\bibnamefont {Pineda}}, \bibinfo
  {author} {\bibfnamefont {D}~\bibnamefont {Midtvedt}}, \ and\ \bibinfo
  {author} {\bibfnamefont {G}~\bibnamefont {Volpe}},\ }\bibfield  {title}
  {\enquote {\bibinfo {title} {Quantitative digital microscopy with deep
  learning},}\ }\href@noop {} {\bibfield  {journal} {\bibinfo  {journal} {arXiv
  preprint arXiv:2010.08260}\ } (\bibinfo {year}
  {2020}{\natexlab{a}})}\BibitemShut {NoStop}%
\bibitem [{\citenamefont {Rivenson}\ \emph
  {et~al.}(2019{\natexlab{a}})\citenamefont {Rivenson}, \citenamefont {Liu},
  \citenamefont {Wei}, \citenamefont {Zhang}, \citenamefont {De~Haan},\ and\
  \citenamefont {Ozcan}}]{rivenson2019phasestain}%
  \BibitemOpen
  \bibfield  {author} {\bibinfo {author} {\bibfnamefont {Y}~\bibnamefont
  {Rivenson}}, \bibinfo {author} {\bibfnamefont {T}~\bibnamefont {Liu}},
  \bibinfo {author} {\bibfnamefont {Z}~\bibnamefont {Wei}}, \bibinfo {author}
  {\bibfnamefont {Y}~\bibnamefont {Zhang}}, \bibinfo {author} {\bibfnamefont
  {K}~\bibnamefont {De~Haan}}, \ and\ \bibinfo {author} {\bibfnamefont
  {A}~\bibnamefont {Ozcan}},\ }\bibfield  {title} {\enquote {\bibinfo {title}
  {Phasestain: the digital staining of label-free quantitative phase microscopy
  images using deep learning},}\ }\href@noop {} {\bibfield  {journal} {\bibinfo
   {journal} {Light Sci. Appl.}\ }\textbf {\bibinfo {volume} {8}},\ \bibinfo
  {pages} {1--11} (\bibinfo {year} {2019}{\natexlab{a}})}\BibitemShut {NoStop}%
\bibitem [{\citenamefont {Zhang}\ \emph {et~al.}(2020)\citenamefont {Zhang},
  \citenamefont {De~Haan}, \citenamefont {Rivenson}, \citenamefont {Li},
  \citenamefont {Delis},\ and\ \citenamefont {Ozcan}}]{zhang2020digital}%
  \BibitemOpen
  \bibfield  {author} {\bibinfo {author} {\bibfnamefont {Y}~\bibnamefont
  {Zhang}}, \bibinfo {author} {\bibfnamefont {K}~\bibnamefont {De~Haan}},
  \bibinfo {author} {\bibfnamefont {Y}~\bibnamefont {Rivenson}}, \bibinfo
  {author} {\bibfnamefont {J}~\bibnamefont {Li}}, \bibinfo {author}
  {\bibfnamefont {A}~\bibnamefont {Delis}}, \ and\ \bibinfo {author}
  {\bibfnamefont {A}~\bibnamefont {Ozcan}},\ }\bibfield  {title} {\enquote
  {\bibinfo {title} {Digital synthesis of histological stains using
  micro-structured and multiplexed virtual staining of label-free tissue},}\
  }\href@noop {} {\bibfield  {journal} {\bibinfo  {journal} {Light Sci. Appl.}\
  }\textbf {\bibinfo {volume} {9}},\ \bibinfo {pages} {1--13} (\bibinfo {year}
  {2020})}\BibitemShut {NoStop}%
\bibitem [{\citenamefont {Rivenson}\ \emph
  {et~al.}(2019{\natexlab{b}})\citenamefont {Rivenson}, \citenamefont {Wang},
  \citenamefont {Wei}, \citenamefont {De~Haan}, \citenamefont {Zhang},
  \citenamefont {Wu}, \citenamefont {G{\"u}nayd{\i}n}, \citenamefont
  {Zuckerman}, \citenamefont {Chong}, \citenamefont {Sisk} \emph
  {et~al.}}]{rivenson2019virtual}%
  \BibitemOpen
  \bibfield  {author} {\bibinfo {author} {\bibfnamefont {Y}~\bibnamefont
  {Rivenson}}, \bibinfo {author} {\bibfnamefont {H}~\bibnamefont {Wang}},
  \bibinfo {author} {\bibfnamefont {Z}~\bibnamefont {Wei}}, \bibinfo {author}
  {\bibfnamefont {K}~\bibnamefont {De~Haan}}, \bibinfo {author} {\bibfnamefont
  {Y}~\bibnamefont {Zhang}}, \bibinfo {author} {\bibfnamefont {Y}~\bibnamefont
  {Wu}}, \bibinfo {author} {\bibfnamefont {H}~\bibnamefont {G{\"u}nayd{\i}n}},
  \bibinfo {author} {\bibfnamefont {J~E}\ \bibnamefont {Zuckerman}}, \bibinfo
  {author} {\bibfnamefont {T}~\bibnamefont {Chong}}, \bibinfo {author}
  {\bibfnamefont {A~E}\ \bibnamefont {Sisk}},  \emph {et~al.},\ }\bibfield
  {title} {\enquote {\bibinfo {title} {Virtual histological staining of
  unlabelled tissue-autofluorescence images via deep learning},}\ }\href@noop
  {} {\bibfield  {journal} {\bibinfo  {journal} {Nat. Biomed. Eng}\ }\textbf
  {\bibinfo {volume} {3}},\ \bibinfo {pages} {466} (\bibinfo {year}
  {2019}{\natexlab{b}})}\BibitemShut {NoStop}%
\bibitem [{\citenamefont {Nygate}\ \emph {et~al.}(2020)\citenamefont {Nygate},
  \citenamefont {Levi}, \citenamefont {Mirsky}, \citenamefont {Turko},
  \citenamefont {Rubin}, \citenamefont {Barnea}, \citenamefont
  {Dardikman-Yoffe}, \citenamefont {Haifler}, \citenamefont {Shalev},\ and\
  \citenamefont {Shaked}}]{nygate2020holographic}%
  \BibitemOpen
  \bibfield  {author} {\bibinfo {author} {\bibfnamefont {Y~N}\ \bibnamefont
  {Nygate}}, \bibinfo {author} {\bibfnamefont {M}~\bibnamefont {Levi}},
  \bibinfo {author} {\bibfnamefont {S~K}\ \bibnamefont {Mirsky}}, \bibinfo
  {author} {\bibfnamefont {N~A}\ \bibnamefont {Turko}}, \bibinfo {author}
  {\bibfnamefont {M}~\bibnamefont {Rubin}}, \bibinfo {author} {\bibfnamefont
  {I}~\bibnamefont {Barnea}}, \bibinfo {author} {\bibfnamefont {G}~\bibnamefont
  {Dardikman-Yoffe}}, \bibinfo {author} {\bibfnamefont {M}~\bibnamefont
  {Haifler}}, \bibinfo {author} {\bibfnamefont {A}~\bibnamefont {Shalev}}, \
  and\ \bibinfo {author} {\bibfnamefont {N~T}\ \bibnamefont {Shaked}},\
  }\bibfield  {title} {\enquote {\bibinfo {title} {Holographic virtual staining
  of individual biological cells},}\ }\href@noop {} {\bibfield  {journal}
  {\bibinfo  {journal} {PNAS}\ }\textbf {\bibinfo {volume} {117}},\ \bibinfo
  {pages} {9223--9231} (\bibinfo {year} {2020})}\BibitemShut {NoStop}%
\bibitem [{\citenamefont {Liu}\ \emph {et~al.}(2020)\citenamefont {Liu},
  \citenamefont {Yuan}, \citenamefont {Wang},\ and\ \citenamefont
  {Ji}}]{liu2020global}%
  \BibitemOpen
  \bibfield  {author} {\bibinfo {author} {\bibfnamefont {Y}~\bibnamefont
  {Liu}}, \bibinfo {author} {\bibfnamefont {H}~\bibnamefont {Yuan}}, \bibinfo
  {author} {\bibfnamefont {Z}~\bibnamefont {Wang}}, \ and\ \bibinfo {author}
  {\bibfnamefont {S}~\bibnamefont {Ji}},\ }\bibfield  {title} {\enquote
  {\bibinfo {title} {Global pixel transformers for virtual staining of
  microscopy images},}\ }\href@noop {} {\bibfield  {journal} {\bibinfo
  {journal} {IEEE Trans Med Imaging}\ }\textbf {\bibinfo {volume} {39}},\
  \bibinfo {pages} {2256--2266} (\bibinfo {year} {2020})}\BibitemShut {NoStop}%
\bibitem [{\citenamefont {Li}\ \emph {et~al.}(2020)\citenamefont {Li},
  \citenamefont {Hui}, \citenamefont {Zhang}, \citenamefont {Tong},
  \citenamefont {Tian}, \citenamefont {Yang}, \citenamefont {Liu},
  \citenamefont {Chen},\ and\ \citenamefont {Tian}}]{li2020deep}%
  \BibitemOpen
  \bibfield  {author} {\bibinfo {author} {\bibfnamefont {D}~\bibnamefont {Li}},
  \bibinfo {author} {\bibfnamefont {H}~\bibnamefont {Hui}}, \bibinfo {author}
  {\bibfnamefont {Y}~\bibnamefont {Zhang}}, \bibinfo {author} {\bibfnamefont
  {W}~\bibnamefont {Tong}}, \bibinfo {author} {\bibfnamefont {F}~\bibnamefont
  {Tian}}, \bibinfo {author} {\bibfnamefont {X}~\bibnamefont {Yang}}, \bibinfo
  {author} {\bibfnamefont {J}~\bibnamefont {Liu}}, \bibinfo {author}
  {\bibfnamefont {Y}~\bibnamefont {Chen}}, \ and\ \bibinfo {author}
  {\bibfnamefont {J}~\bibnamefont {Tian}},\ }\bibfield  {title} {\enquote
  {\bibinfo {title} {Deep learning for virtual histological staining of
  bright-field microscopic images of unlabeled carotid artery tissue.}}\
  }\href@noop {} {\bibfield  {journal} {\bibinfo  {journal} {Mol. Imaging.
  Biol.}\ }\textbf {\bibinfo {volume} {22}},\ \bibinfo {pages} {1301--1309}
  (\bibinfo {year} {2020})}\BibitemShut {NoStop}%
\bibitem [{\citenamefont {Buzzetti}\ \emph {et~al.}(2016)\citenamefont
  {Buzzetti}, \citenamefont {Pinzani},\ and\ \citenamefont
  {Tsochatzis}}]{buzzetti2016multiple}%
  \BibitemOpen
  \bibfield  {author} {\bibinfo {author} {\bibfnamefont {E}~\bibnamefont
  {Buzzetti}}, \bibinfo {author} {\bibfnamefont {M}~\bibnamefont {Pinzani}}, \
  and\ \bibinfo {author} {\bibfnamefont {E~A}\ \bibnamefont {Tsochatzis}},\
  }\bibfield  {title} {\enquote {\bibinfo {title} {The multiple-hit
  pathogenesis of non-alcoholic fatty liver disease ({NAFLD})},}\ }\href@noop
  {} {\bibfield  {journal} {\bibinfo  {journal} {Metabolism}\ }\textbf
  {\bibinfo {volume} {65}},\ \bibinfo {pages} {1038--1048} (\bibinfo {year}
  {2016})}\BibitemShut {NoStop}%
\bibitem [{\citenamefont {Blakney}\ \emph {et~al.}(2020)\citenamefont
  {Blakney}, \citenamefont {Deletic}, \citenamefont {McKay}, \citenamefont
  {Bouton}, \citenamefont {Ashford}, \citenamefont {Shattock},\ and\
  \citenamefont {Sabirsh}}]{blakney2020effect}%
  \BibitemOpen
  \bibfield  {author} {\bibinfo {author} {\bibfnamefont {A~K}\ \bibnamefont
  {Blakney}}, \bibinfo {author} {\bibfnamefont {P}~\bibnamefont {Deletic}},
  \bibinfo {author} {\bibfnamefont {P~F}\ \bibnamefont {McKay}}, \bibinfo
  {author} {\bibfnamefont {C~R}\ \bibnamefont {Bouton}}, \bibinfo {author}
  {\bibfnamefont {M}~\bibnamefont {Ashford}}, \bibinfo {author} {\bibfnamefont
  {R~J}\ \bibnamefont {Shattock}}, \ and\ \bibinfo {author} {\bibfnamefont
  {A}~\bibnamefont {Sabirsh}},\ }\bibfield  {title} {\enquote {\bibinfo {title}
  {Effect of complexing lipids on cellular uptake and expression of messenger
  {RNA} in human skin explants},}\ }\href@noop {} {\bibfield  {journal}
  {\bibinfo  {journal} {J. Control. Release}\ ,\ \bibinfo {pages}
  {doi:10.1016/j.jconrel.2020.11.033}} (\bibinfo {year} {2020})}\BibitemShut
  {NoStop}%
\bibitem [{\citenamefont {Bartesaghi}\ \emph {et~al.}(2015)\citenamefont
  {Bartesaghi}, \citenamefont {Hallen}, \citenamefont {Huang}, \citenamefont
  {Svensson}, \citenamefont {Momo}, \citenamefont {Wallin}, \citenamefont
  {Carlsson}, \citenamefont {Forsl{\"o}w}, \citenamefont {Seale},\ and\
  \citenamefont {Peng}}]{bartesaghi2015thermogenic}%
  \BibitemOpen
  \bibfield  {author} {\bibinfo {author} {\bibfnamefont {S}~\bibnamefont
  {Bartesaghi}}, \bibinfo {author} {\bibfnamefont {S}~\bibnamefont {Hallen}},
  \bibinfo {author} {\bibfnamefont {L}~\bibnamefont {Huang}}, \bibinfo {author}
  {\bibfnamefont {P~A}\ \bibnamefont {Svensson}}, \bibinfo {author}
  {\bibfnamefont {R~A}\ \bibnamefont {Momo}}, \bibinfo {author} {\bibfnamefont
  {S}~\bibnamefont {Wallin}}, \bibinfo {author} {\bibfnamefont {E~K}\
  \bibnamefont {Carlsson}}, \bibinfo {author} {\bibfnamefont {A}~\bibnamefont
  {Forsl{\"o}w}}, \bibinfo {author} {\bibfnamefont {P}~\bibnamefont {Seale}}, \
  and\ \bibinfo {author} {\bibfnamefont {X~R}\ \bibnamefont {Peng}},\
  }\bibfield  {title} {\enquote {\bibinfo {title} {Thermogenic activity of
  {UCP}1 in human white fat-derived beige adipocytes},}\ }\href@noop {}
  {\bibfield  {journal} {\bibinfo  {journal} {J. Mol. Endocrinol.}\ }\textbf
  {\bibinfo {volume} {29}},\ \bibinfo {pages} {130--139} (\bibinfo {year}
  {2015})}\BibitemShut {NoStop}%
\bibitem [{\citenamefont {McQuin}\ \emph {et~al.}(2018)\citenamefont {McQuin},
  \citenamefont {Goodman}, \citenamefont {Chernyshev}, \citenamefont
  {Kamentsky}, \citenamefont {Cimini}, \citenamefont {Karhohs}, \citenamefont
  {Doan}, \citenamefont {Ding}, \citenamefont {Rafelski}, \citenamefont
  {Thirstrup} \emph {et~al.}}]{mcquin2018cellprofiler}%
  \BibitemOpen
  \bibfield  {author} {\bibinfo {author} {\bibfnamefont {C}~\bibnamefont
  {McQuin}}, \bibinfo {author} {\bibfnamefont {A}~\bibnamefont {Goodman}},
  \bibinfo {author} {\bibfnamefont {V}~\bibnamefont {Chernyshev}}, \bibinfo
  {author} {\bibfnamefont {L}~\bibnamefont {Kamentsky}}, \bibinfo {author}
  {\bibfnamefont {B~A}\ \bibnamefont {Cimini}}, \bibinfo {author}
  {\bibfnamefont {K~W}\ \bibnamefont {Karhohs}}, \bibinfo {author}
  {\bibfnamefont {M}~\bibnamefont {Doan}}, \bibinfo {author} {\bibfnamefont
  {L}~\bibnamefont {Ding}}, \bibinfo {author} {\bibfnamefont {S~M}\
  \bibnamefont {Rafelski}}, \bibinfo {author} {\bibfnamefont {D}~\bibnamefont
  {Thirstrup}},  \emph {et~al.},\ }\bibfield  {title} {\enquote {\bibinfo
  {title} {Cellprofiler 3.0: Next-generation image processing for biology},}\
  }\href@noop {} {\bibfield  {journal} {\bibinfo  {journal} {PLoS Biol.}\
  }\textbf {\bibinfo {volume} {16}},\ \bibinfo {pages} {e2005970} (\bibinfo
  {year} {2018})}\BibitemShut {NoStop}%
\bibitem [{\citenamefont {Midtvedt}\ \emph
  {et~al.}(2020{\natexlab{b}})\citenamefont {Midtvedt}, \citenamefont {Pineda},
  \citenamefont {Helgadottir}, \citenamefont {Midtvedt},\ and\ \citenamefont
  {Volpe}}]{VirtualStainingRepo}%
  \BibitemOpen
  \bibfield  {author} {\bibinfo {author} {\bibfnamefont {B}~\bibnamefont
  {Midtvedt}}, \bibinfo {author} {\bibfnamefont {J}~\bibnamefont {Pineda}},
  \bibinfo {author} {\bibfnamefont {S}~\bibnamefont {Helgadottir}}, \bibinfo
  {author} {\bibfnamefont {D}~\bibnamefont {Midtvedt}}, \ and\ \bibinfo
  {author} {\bibfnamefont {G}~\bibnamefont {Volpe}},\ }\href@noop {} {\enquote
  {\bibinfo {title} {{VirtualStaining}},}\ }\bibinfo {howpublished}
  {\url{https://github.com/softmatterlab/VirtualStaining}} (\bibinfo {year}
  {2020}{\natexlab{b}})\BibitemShut {NoStop}%
\bibitem [{\citenamefont {Mehlig}(2019)}]{mehlig2019artificial}%
  \BibitemOpen
  \bibfield  {author} {\bibinfo {author} {\bibfnamefont {B}~\bibnamefont
  {Mehlig}},\ }\bibfield  {title} {\enquote {\bibinfo {title} {Artificial
  neural networks},}\ }\href@noop {} {\bibfield  {journal} {\bibinfo  {journal}
  {arXiv preprint arXiv:1901.05639}\ } (\bibinfo {year} {2019})}\BibitemShut
  {NoStop}%
\bibitem [{\citenamefont {Goodfellow}\ \emph {et~al.}(2014)\citenamefont
  {Goodfellow}, \citenamefont {Pouget-Abadie}, \citenamefont {Mirza},
  \citenamefont {Xu}, \citenamefont {Warde-Farley}, \citenamefont {Ozair},
  \citenamefont {Courville},\ and\ \citenamefont {Bengio}}]{NIPS2014_5ca3e9b1}%
  \BibitemOpen
  \bibfield  {author} {\bibinfo {author} {\bibfnamefont {I}~\bibnamefont
  {Goodfellow}}, \bibinfo {author} {\bibfnamefont {J}~\bibnamefont
  {Pouget-Abadie}}, \bibinfo {author} {\bibfnamefont {M}~\bibnamefont {Mirza}},
  \bibinfo {author} {\bibfnamefont {B}~\bibnamefont {Xu}}, \bibinfo {author}
  {\bibfnamefont {D}~\bibnamefont {Warde-Farley}}, \bibinfo {author}
  {\bibfnamefont {S}~\bibnamefont {Ozair}}, \bibinfo {author} {\bibfnamefont
  {A}~\bibnamefont {Courville}}, \ and\ \bibinfo {author} {\bibfnamefont
  {Y}~\bibnamefont {Bengio}},\ }\bibfield  {title} {\enquote {\bibinfo {title}
  {Generative adversarial nets},}\ }in\ \href@noop {} {\emph {\bibinfo
  {booktitle} {Advances in Neural Information Processing Systems}}},\
  Vol.~\bibinfo {volume} {27}\ (\bibinfo {year} {2014})\ pp.\ \bibinfo {pages}
  {2672--2680}\BibitemShut {NoStop}%
\bibitem [{\citenamefont {He}\ \emph {et~al.}(2016)\citenamefont {He},
  \citenamefont {Zhang}, \citenamefont {Ren},\ and\ \citenamefont
  {Sun}}]{he2016deep}%
  \BibitemOpen
  \bibfield  {author} {\bibinfo {author} {\bibfnamefont {K}~\bibnamefont {He}},
  \bibinfo {author} {\bibfnamefont {X}~\bibnamefont {Zhang}}, \bibinfo {author}
  {\bibfnamefont {S}~\bibnamefont {Ren}}, \ and\ \bibinfo {author}
  {\bibfnamefont {J}~\bibnamefont {Sun}},\ }\bibfield  {title} {\enquote
  {\bibinfo {title} {Deep residual learning for image recognition},}\ }in\
  \href@noop {} {\emph {\bibinfo {booktitle} {Proc IEEE Int Conf Comput Vis}}}\
  (\bibinfo {year} {2016})\ pp.\ \bibinfo {pages} {770--778}\BibitemShut
  {NoStop}%
\bibitem [{\citenamefont {Isola}\ \emph {et~al.}(2017)\citenamefont {Isola},
  \citenamefont {Zhu}, \citenamefont {Zhou},\ and\ \citenamefont
  {Efros}}]{isola2017image}%
  \BibitemOpen
  \bibfield  {author} {\bibinfo {author} {\bibfnamefont {P}~\bibnamefont
  {Isola}}, \bibinfo {author} {\bibfnamefont {J~Y}\ \bibnamefont {Zhu}},
  \bibinfo {author} {\bibfnamefont {T}~\bibnamefont {Zhou}}, \ and\ \bibinfo
  {author} {\bibfnamefont {A~A}\ \bibnamefont {Efros}},\ }\bibfield  {title}
  {\enquote {\bibinfo {title} {Image-to-image translation with conditional
  adversarial networks},}\ }in\ \href@noop {} {\emph {\bibinfo {booktitle}
  {Proc IEEE Int Conf Comput Vis}}}\ (\bibinfo {year} {2017})\ pp.\ \bibinfo
  {pages} {1125--1134}\BibitemShut {NoStop}%
\bibitem [{\citenamefont {Mirza}\ and\ \citenamefont
  {Osindero}(2014)}]{mirza2014conditional}%
  \BibitemOpen
  \bibfield  {author} {\bibinfo {author} {\bibfnamefont {M}~\bibnamefont
  {Mirza}}\ and\ \bibinfo {author} {\bibfnamefont {S}~\bibnamefont
  {Osindero}},\ }\bibfield  {title} {\enquote {\bibinfo {title} {Conditional
  generative adversarial nets},}\ }\href@noop {} {\bibfield  {journal}
  {\bibinfo  {journal} {arXiv preprint arXiv:1411.1784}\ } (\bibinfo {year}
  {2014})}\BibitemShut {NoStop}%
\bibitem [{\citenamefont {Ronneberger}\ \emph {et~al.}(2015)\citenamefont
  {Ronneberger}, \citenamefont {Fischer},\ and\ \citenamefont
  {Brox}}]{ronneberger2015u}%
  \BibitemOpen
  \bibfield  {author} {\bibinfo {author} {\bibfnamefont {O}~\bibnamefont
  {Ronneberger}}, \bibinfo {author} {\bibfnamefont {P}~\bibnamefont {Fischer}},
  \ and\ \bibinfo {author} {\bibfnamefont {T}~\bibnamefont {Brox}},\ }\bibfield
   {title} {\enquote {\bibinfo {title} {U-net: Convolutional networks for
  biomedical image segmentation},}\ }in\ \href@noop {} {\emph {\bibinfo
  {booktitle} {Int Conf Med Image Comput Comput Assist}}}\ (\bibinfo
  {organization} {Springer},\ \bibinfo {year} {2015})\ pp.\ \bibinfo {pages}
  {234--241}\BibitemShut {NoStop}%
\bibitem [{\citenamefont {Lei}\ \emph {et~al.}(2020)\citenamefont {Lei},
  \citenamefont {Li}, \citenamefont {Zhang},\ and\ \citenamefont
  {Li}}]{lei2020wavelet}%
  \BibitemOpen
  \bibfield  {author} {\bibinfo {author} {\bibfnamefont {Y}~\bibnamefont
  {Lei}}, \bibinfo {author} {\bibfnamefont {D}~\bibnamefont {Li}}, \bibinfo
  {author} {\bibfnamefont {H}~\bibnamefont {Zhang}}, \ and\ \bibinfo {author}
  {\bibfnamefont {X}~\bibnamefont {Li}},\ }\bibfield  {title} {\enquote
  {\bibinfo {title} {Wavelet feature outdoor fingerprint localization based on
  resnet and deep convolution gan},}\ }\href@noop {} {\bibfield  {journal}
  {\bibinfo  {journal} {Symmetry}\ }\textbf {\bibinfo {volume} {12}},\ \bibinfo
  {pages} {1565} (\bibinfo {year} {2020})}\BibitemShut {NoStop}%
\bibitem [{\citenamefont {Xu}\ \emph {et~al.}(2015)\citenamefont {Xu},
  \citenamefont {Wang}, \citenamefont {Chen},\ and\ \citenamefont
  {Li}}]{xu2015empirical}%
  \BibitemOpen
  \bibfield  {author} {\bibinfo {author} {\bibfnamefont {B}~\bibnamefont {Xu}},
  \bibinfo {author} {\bibfnamefont {N}~\bibnamefont {Wang}}, \bibinfo {author}
  {\bibfnamefont {T}~\bibnamefont {Chen}}, \ and\ \bibinfo {author}
  {\bibfnamefont {M}~\bibnamefont {Li}},\ }\bibfield  {title} {\enquote
  {\bibinfo {title} {Empirical evaluation of rectified activations in
  convolutional network},}\ }\href@noop {} {\bibfield  {journal} {\bibinfo
  {journal} {arXiv preprint arXiv:1505.00853}\ } (\bibinfo {year}
  {2015})}\BibitemShut {NoStop}%
\bibitem [{\citenamefont {Foster}(2019)}]{foster2019generative}%
  \BibitemOpen
  \bibfield  {author} {\bibinfo {author} {\bibfnamefont {D}~\bibnamefont
  {Foster}},\ }\href@noop {} {\emph {\bibinfo {title} {Generative deep
  learning: teaching machines to paint, write, compose, and play}}}\ (\bibinfo
  {publisher} {O'Reilly Media},\ \bibinfo {year} {2019})\BibitemShut {NoStop}%
\bibitem [{\citenamefont {Midtvedt}\ \emph
  {et~al.}(2020{\natexlab{c}})\citenamefont {Midtvedt}, \citenamefont
  {Helgadottir}, \citenamefont {Argun}, \citenamefont {Pineda}, \citenamefont
  {Midtvedt},\ and\ \citenamefont {Volpe}}]{DeepTrack}%
  \BibitemOpen
  \bibfield  {author} {\bibinfo {author} {\bibfnamefont {B}~\bibnamefont
  {Midtvedt}}, \bibinfo {author} {\bibfnamefont {S}~\bibnamefont
  {Helgadottir}}, \bibinfo {author} {\bibfnamefont {A}~\bibnamefont {Argun}},
  \bibinfo {author} {\bibfnamefont {J}~\bibnamefont {Pineda}}, \bibinfo
  {author} {\bibfnamefont {D}~\bibnamefont {Midtvedt}}, \ and\ \bibinfo
  {author} {\bibfnamefont {G}~\bibnamefont {Volpe}},\ }\href@noop {} {\enquote
  {\bibinfo {title} {Deeptrack 2.0},}\ }\bibinfo {howpublished}
  {\url{https://github.com/softmatterlab/DeepTrack-2.0}} (\bibinfo {year}
  {2020}{\natexlab{c}})\BibitemShut {NoStop}%
\bibitem [{\citenamefont {Chollet}\ \emph {et~al.}(2015)\citenamefont {Chollet}
  \emph {et~al.}}]{chollet2015keras}%
  \BibitemOpen
  \bibfield  {author} {\bibinfo {author} {\bibfnamefont {F}~\bibnamefont
  {Chollet}} \emph {et~al.},\ }\href {https://keras.io.} {\enquote {\bibinfo
  {title} {Keras},}\ } (\bibinfo {year} {2015})\BibitemShut {NoStop}%
\bibitem [{\citenamefont {Abadi}\ \emph {et~al.}(2016)\citenamefont {Abadi},
  \citenamefont {Barham}, \citenamefont {Chen}, \citenamefont {Chen},
  \citenamefont {Davis}, \citenamefont {Dean}, \citenamefont {Devin},
  \citenamefont {Ghemawat}, \citenamefont {Irving}, \citenamefont {Isard} \emph
  {et~al.}}]{abadi2016tensorflow}%
  \BibitemOpen
  \bibfield  {author} {\bibinfo {author} {\bibfnamefont {M}~\bibnamefont
  {Abadi}}, \bibinfo {author} {\bibfnamefont {P}~\bibnamefont {Barham}},
  \bibinfo {author} {\bibfnamefont {J}~\bibnamefont {Chen}}, \bibinfo {author}
  {\bibfnamefont {Z}~\bibnamefont {Chen}}, \bibinfo {author} {\bibfnamefont
  {A}~\bibnamefont {Davis}}, \bibinfo {author} {\bibfnamefont {J}~\bibnamefont
  {Dean}}, \bibinfo {author} {\bibfnamefont {M}~\bibnamefont {Devin}}, \bibinfo
  {author} {\bibfnamefont {S}~\bibnamefont {Ghemawat}}, \bibinfo {author}
  {\bibfnamefont {G}~\bibnamefont {Irving}}, \bibinfo {author} {\bibfnamefont
  {M}~\bibnamefont {Isard}},  \emph {et~al.},\ }\bibfield  {title} {\enquote
  {\bibinfo {title} {Tensorflow: A system for large-scale machine learning},}\
  }in\ \href@noop {} {\emph {\bibinfo {booktitle} {12th {USENIX} symposium on
  operating systems design and implementation ({OSDI} 16)}}}\ (\bibinfo {year}
  {2016})\ pp.\ \bibinfo {pages} {265--283}\BibitemShut {NoStop}%
\bibitem [{\citenamefont {McClelland}\ \emph {et~al.}(1986)\citenamefont
  {McClelland}, \citenamefont {Rumelhart}, \citenamefont {Group} \emph
  {et~al.}}]{mcclelland1986parallel}%
  \BibitemOpen
  \bibfield  {author} {\bibinfo {author} {\bibfnamefont {J~L}\ \bibnamefont
  {McClelland}}, \bibinfo {author} {\bibfnamefont {D~E}\ \bibnamefont
  {Rumelhart}}, \bibinfo {author} {\bibfnamefont {PDP~Research}\ \bibnamefont
  {Group}},  \emph {et~al.},\ }\bibfield  {title} {\enquote {\bibinfo {title}
  {Parallel distributed processing},}\ }\href@noop {} {\bibfield  {journal}
  {\bibinfo  {journal} {Explor. Microstructure Cognition}\ }\textbf {\bibinfo
  {volume} {2}},\ \bibinfo {pages} {216--271} (\bibinfo {year}
  {1986})}\BibitemShut {NoStop}%
\bibitem [{\citenamefont {Kingma}\ and\ \citenamefont
  {Ba}(2014)}]{kingma2014adam}%
  \BibitemOpen
  \bibfield  {author} {\bibinfo {author} {\bibfnamefont {D~P}\ \bibnamefont
  {Kingma}}\ and\ \bibinfo {author} {\bibfnamefont {J}~\bibnamefont {Ba}},\
  }\bibfield  {title} {\enquote {\bibinfo {title} {Adam: A method for
  stochastic optimization},}\ }\href@noop {} {\bibfield  {journal} {\bibinfo
  {journal} {arXiv preprint arXiv:1412.6980}\ } (\bibinfo {year}
  {2014})}\BibitemShut {NoStop}%
\bibitem [{\citenamefont {Yanina}\ \emph {et~al.}(2018)\citenamefont {Yanina},
  \citenamefont {Lazareva},\ and\ \citenamefont {Tuchin}}]{Yanina:18}%
  \BibitemOpen
  \bibfield  {author} {\bibinfo {author} {\bibfnamefont {I~Y}\ \bibnamefont
  {Yanina}}, \bibinfo {author} {\bibfnamefont {E~N}\ \bibnamefont {Lazareva}},
  \ and\ \bibinfo {author} {\bibfnamefont {V~V}\ \bibnamefont {Tuchin}},\
  }\bibfield  {title} {\enquote {\bibinfo {title} {Refractive index of adipose
  tissue and lipid droplet measured in wide spectral and temperature ranges},}\
  }\href@noop {} {\bibfield  {journal} {\bibinfo  {journal} {Appl. Opt.}\
  }\textbf {\bibinfo {volume} {57}},\ \bibinfo {pages} {4839--4848} (\bibinfo
  {year} {2018})}\BibitemShut {NoStop}%
\bibitem [{\citenamefont {Robenek}\ \emph {et~al.}(2004)\citenamefont
  {Robenek}, \citenamefont {Severs}, \citenamefont {Schlattmann}, \citenamefont
  {Plenz}, \citenamefont {Zimmer}, \citenamefont {Troyer},\ and\ \citenamefont
  {Robenek}}]{robenek2004lipids}%
  \BibitemOpen
  \bibfield  {author} {\bibinfo {author} {\bibfnamefont {M~J}\ \bibnamefont
  {Robenek}}, \bibinfo {author} {\bibfnamefont {N~J}\ \bibnamefont {Severs}},
  \bibinfo {author} {\bibfnamefont {K}~\bibnamefont {Schlattmann}}, \bibinfo
  {author} {\bibfnamefont {G}~\bibnamefont {Plenz}}, \bibinfo {author}
  {\bibfnamefont {K~P}\ \bibnamefont {Zimmer}}, \bibinfo {author}
  {\bibfnamefont {D}~\bibnamefont {Troyer}}, \ and\ \bibinfo {author}
  {\bibfnamefont {H}~\bibnamefont {Robenek}},\ }\bibfield  {title} {\enquote
  {\bibinfo {title} {Lipids partition caveolin-1 from er membranes into lipid
  droplets: updating the model of lipid droplet biogenesis},}\ }\href@noop {}
  {\bibfield  {journal} {\bibinfo  {journal} {FASEB J.}\ }\textbf {\bibinfo
  {volume} {18}},\ \bibinfo {pages} {866--868} (\bibinfo {year}
  {2004})}\BibitemShut {NoStop}%
\bibitem [{\citenamefont {Sch{\"{u}}rmann}\ \emph {et~al.}(2016)\citenamefont
  {Sch{\"{u}}rmann}, \citenamefont {Scholze}, \citenamefont {M{\"{u}}ller},
  \citenamefont {Guck},\ and\ \citenamefont {Chan}}]{Schurmann2016}%
  \BibitemOpen
  \bibfield  {author} {\bibinfo {author} {\bibfnamefont {M}~\bibnamefont
  {Sch{\"{u}}rmann}}, \bibinfo {author} {\bibfnamefont {J}~\bibnamefont
  {Scholze}}, \bibinfo {author} {\bibfnamefont {P}~\bibnamefont
  {M{\"{u}}ller}}, \bibinfo {author} {\bibfnamefont {J}~\bibnamefont {Guck}}, \
  and\ \bibinfo {author} {\bibfnamefont {C~J.}\ \bibnamefont {Chan}},\
  }\bibfield  {title} {\enquote {\bibinfo {title} {{Cell nuclei have lower
  refractive index and mass density than cytoplasm}},}\ }\href {\doibase
  10.1002/jbio.201500273} {\bibfield  {journal} {\bibinfo  {journal} {J.
  Biophoton.}\ }\textbf {\bibinfo {volume} {9}},\ \bibinfo {pages} {1068--1076}
  (\bibinfo {year} {2016})}\BibitemShut {NoStop}%
\bibitem [{\citenamefont {Van~Herpen}\ and\ \citenamefont
  {Schrauwen-Hinderling}(2008)}]{van2008lipid}%
  \BibitemOpen
  \bibfield  {author} {\bibinfo {author} {\bibfnamefont {N~A}\ \bibnamefont
  {Van~Herpen}}\ and\ \bibinfo {author} {\bibfnamefont {V~B}\ \bibnamefont
  {Schrauwen-Hinderling}},\ }\bibfield  {title} {\enquote {\bibinfo {title}
  {Lipid accumulation in non-adipose tissue and lipotoxicity},}\ }\href@noop {}
  {\bibfield  {journal} {\bibinfo  {journal} {Physiol. Behav.}\ }\textbf
  {\bibinfo {volume} {94}},\ \bibinfo {pages} {231--241} (\bibinfo {year}
  {2008})}\BibitemShut {NoStop}%
\bibitem [{\citenamefont {Charras}(2008)}]{charras2008short}%
  \BibitemOpen
  \bibfield  {author} {\bibinfo {author} {\bibfnamefont {G~T}\ \bibnamefont
  {Charras}},\ }\bibfield  {title} {\enquote {\bibinfo {title} {A short history
  of blebbing},}\ }\href@noop {} {\bibfield  {journal} {\bibinfo  {journal} {J.
  Microsc.}\ }\textbf {\bibinfo {volume} {231}},\ \bibinfo {pages} {466--478}
  (\bibinfo {year} {2008})}\BibitemShut {NoStop}%
\bibitem [{\citenamefont {Purschke}\ \emph {et~al.}(2010)\citenamefont
  {Purschke}, \citenamefont {Rubio}, \citenamefont {Held},\ and\ \citenamefont
  {Redmond}}]{purschke2010phototoxicity}%
  \BibitemOpen
  \bibfield  {author} {\bibinfo {author} {\bibfnamefont {M}~\bibnamefont
  {Purschke}}, \bibinfo {author} {\bibfnamefont {N}~\bibnamefont {Rubio}},
  \bibinfo {author} {\bibfnamefont {K~D}\ \bibnamefont {Held}}, \ and\ \bibinfo
  {author} {\bibfnamefont {R~W}\ \bibnamefont {Redmond}},\ }\bibfield  {title}
  {\enquote {\bibinfo {title} {Phototoxicity of hoechst 33342 in time-lapse
  fluorescence microscopy},}\ }\href@noop {} {\bibfield  {journal} {\bibinfo
  {journal} {Photochem. Photobiol. Sci.}\ }\textbf {\bibinfo {volume} {9}},\
  \bibinfo {pages} {1634--1639} (\bibinfo {year} {2010})}\BibitemShut {NoStop}%
\end{thebibliography}


\end{document}